\documentclass[fleqn,usenatbib]{mnras}

\usepackage{newtxtext,newtxmath}

\usepackage[T1]{fontenc}

\DeclareRobustCommand{\VAN}[3]{#2}
\let\VANthebibliography\thebibliography
\def\thebibliography{\DeclareRobustCommand{\VAN}[3]{##3}\VANthebibliography}

\usepackage{graphicx}	
\usepackage{amsmath}	
\usepackage{orcidlink}
\newcommand{\superstars}{\textsc{Superstars }}
\newcommand{\auriga}{\textsc{auriga }}
\usepackage{xcolor}

\title[Auriga Superstars: spiral arms]{The diverse nature of spiral arms in the Auriga Superstars cosmological hydrodynamic simulations}

\author[R. J. J. Grand et al.]{Robert J.~J.~Grand$^1$\thanks{r.j.grand@ljmu.ac.uk}\orcidlink{0000-0001-9667-1340},
Francesca Fragkoudi$^{2}$\orcidlink{0000-0002-0897-3013},
R\"udiger Pakmor$^{3}$\orcidlink{0000-0003-3308-2420},
Facundo A.~G\'{o}mez$^{4}$\orcidlink{0000-0003-4232-8584},
\newauthor
Freeke van~de~Voort$^5$\orcidlink{0000-0002-6301-638X},
Rebekka Bieri$^6$\orcidlink{0000-0002-4554-4488},
Sophie Townson$^1$
\vspace*{0.1cm}\\
$^{1}$Astrophysics Research Institute, Liverpool John Moores University, 146 Brownlow Hill, Liverpool, L3 5RF, UK\\
$^{2}$Institute for Computational Cosmology, Department of Physics, Durham University, South Road, Durham DH1 3LE, UK \\
$^{3}$Max-Planck-Institut f\"{u}r Astrophysik, Karl-Schwarzschild-Str. 1, D-85748, Garching, Germany\\
$^{4}$Departamento de Astronom\'{i}a, Universidad de La Serena, Av.~Juan Cisternas 1200 Norte, La Serena, Chile\\
$^{5}$Cardiff Hub for Astrophysics Research and Technology, School of Physics and Astronomy, Cardiff University, Queen’s Buildings, Cardiff CF24 3AA, UK\\
$^{6}$Institut für Astrophysik, Universität Zürich, Winterthurerstrasse 190, 8057 Zürich, Switzerland\\
}

\date{Accepted XXX. Received YYY; in original form ZZZ}

\pubyear{\the\year{}}

\begin{document}
\label{firstpage}
\pagerange{\pageref{firstpage}--\pageref{lastpage}}
\maketitle

\begin{abstract}
The dynamical nature and formation mechanism(s) of galactic spiral arms remain long-standing problems in astrophysics. Most theoretical work is based on analytic calculations or idealised simulations, which has yielded several theories of spiral structure. The radial profile of the spiral arm rotation speed - the pattern speed - is a key observable prediction of these theories. However, observations that infer spiral pattern speeds reveal a mixed picture with no clear consensus. Here, we expand on theoretical efforts by examining the pattern speed profiles in the \auriga \superstars set of high-resolution cosmological magnetohydrodynamic simulatons of Milky Way-mass spiral disc galaxies. These simulations combine galaxy formation in a cosmological environment with the high dynamical fidelity afforded by an $\sim 800$ $\rm M_{\odot}$ star particle resolution, giving $\sim 100$ million star particles in the disc. We show that several different spiral arm theories are realised among our simulations, including large-scale kinematic density waves, manifold spirals, dynamic (co-rotating) spirals, and overlapping modes. In particular, we demonstrate that a strong tidal interaction leads to clear kinematic density waves, and that manifold spirals are present in a strongly-barred galaxy. Interestingly, we find that the same galaxy may show qualitative evolution of their spiral pattern speed profiles, indicating that the nature of spiral arms can evolve on potentially sub-Gigayear timescales. Our results demonstrate that in the absence of a strong external encounter or a strong bar, galactic spiral structure is highly transitional and complex with no clear long-lived underlying wave.
\end{abstract}




\section{Introduction}

Galactic spiral arms are the defining features of most star-forming disc galaxies observed both in the local Universe \citep[e.g.][]{BVS05,Elmegreen2011,Leroy2021}, including the Milky Way \citep{BHG16}, and as far away as redshift $\gtrsim 2$ \citep[e.g.][]{Elmegreen2014,Tsukui2024}. Spirals are important for disc galaxy evolution in at least two ways: first, they are sites of ongoing star formation \citep[e.g.][]{Vogel1998,Querejeta2021}; and second, they redistribute angular momentum of galactic material \citep{LBK72,S14R}, thereby driving the radial migration of stars and gas \citep[e.g.][]{SB02} that shapes the stellar chronochemodynamic structure of our own Galaxy \citep[e.g.][]{CSA11,HBH15,Kawata2018,Kawata2024}. Precisely how spiral arms affect galaxy evolution depends on their underlying nature, which is a rich dynamical problem itself \citep[see e.g.][]{Dobbs2014,Hamilton2025}. Despite the prevalence, importance, and beauty of spiral arms, their physical nature remains the subject of a long-standing debate in modern astrophysics with a wide variety of proposed mechanisms.

\subsection{Predictions from theory -- Spiral pattern speeds}

\begin{figure*}
\centering
\includegraphics[scale=1,trim={0 0 0 0}, clip]{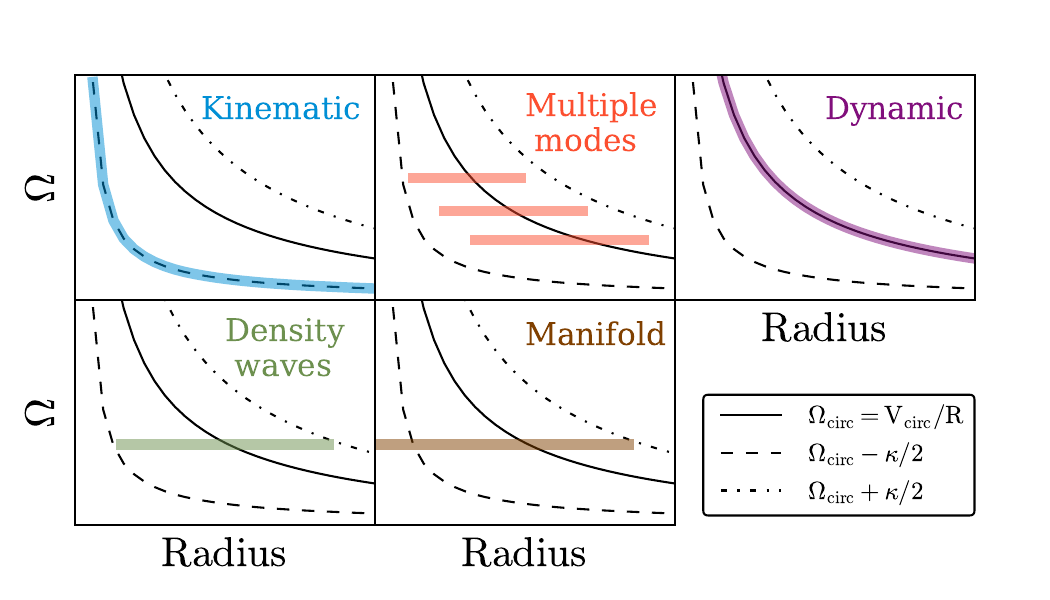}
    \caption{Schematic diagrams of spiral pattern speed radial profiles associated with various prominent spiral arm theories, including: kinematic density wave theory (top-left panel); classic density wave theory (bottom-left panel); multiple mode theory (top-middle panel); manifold theory (bottom-middle panel); and dynamic/co-rotating spiral arms (top-right panel). Each profile is depicted with a shaded band and has a qualitatively unique form.}
    \label{fig:schem}
\end{figure*}

An important property of spiral arms is the speed at which they rotate around the galaxy, which is conventionally referred to as the spiral pattern speed. The form of the spiral pattern speed radial profile is a key distinguishing predictor for theories of spiral structure: in particular, how it relates to the circular angular velocity curve, $\Omega _{\rm circ}$, and the inner and outer Lindblad resonances defined by $\Omega _{\rm circ} \pm \kappa/m$, where $\kappa$ is the radial epicyclic frequency (the frequency at which a star oscillates radially around its average circular orbit) and $m$ is the number of spiral arms in the pattern. Although the radial pattern speed diagnostic may simplify the true complexity of spiral arms, it represents a clear first step to understand their nature.

One of the earliest descriptions of spiral structure is {\bf \emph{kinematic density waves}} \citep[see e.g.][]{L60,Lindblad1963,Kalnajs1973}. The top-left panel of Fig.~\ref{fig:schem} shows the pattern speed profile for kinematic density waves, in which spiral arms have a pattern speed, $\Omega _p = \Omega _{\rm circ} -\kappa/m$, i.e, following the inner Lindblad resonance (ILR) curve for a range of radii. In a frame rotating at $\Omega _p$, stellar orbits appear as closed ellipses, which can be arranged such that orbits align to form a spiral shape. The precise radial profile of the pattern speed is therefore set by the gravitational potential: it can decrease steeply in the inner disc but plateau to an almost constant profile in the outer disc. Thus, kinematic density waves\footnote{The term ``kinematic density waves'' is a misnomer in the sense that they are not actual waves that propagate through the disc at some group velocity \citep[see e.g.][for more details]{Dobbs2014}.} can provide sustained spiral structure that would slowly wind-up on timescales longer than the dynamical timescale. Kinematic density waves may be expected to occur in response to strong tidal interactions \citep[e.g.][]{Donner1994} and in regimes where self-gravity is unimportant: \citet{HS15} found that a weakly self-gravitating $N$-body disc formed spirals whose pattern speeds traced the ILR. 

Inspired by the kinematic density waves of \citet{Lindblad1963}, \citet{LS64} proposed the Quasi-Stationary Spiral Structure (QSSS) hypothesis. At the heart of this theory and its many subsequent developments \citep[e.g.][]{Kalnajs1965,Mark1974,BL96} is the assumption that spiral arms are not material in nature but are instead the crest of a long-lived wave that follows a dispersion relation\footnote{Strictly speaking, the Lin-Shu theory is a linear theory derived under the ``tight-winding'' approximation, i.e., the short wavelength branch of the dispersion relation.}. The bottom-left panel of Fig.~\ref{fig:schem} shows the constant pattern speed of spiral structure proposed by classic {\bf \emph{density wave theory}}, which spans a radial range bounded by the inner Lindblad resonance and the outer Lindblad resonance \citep{LBK72}. The constant pattern speed implies the rigid rotation of spiral structure around the galactic centre, which allows it to preserve its shape over many dynamical times. There is a single co-rotation resonance, defined as the radius at which the spiral pattern speed and galactic rotation curve are equal; stars and gas rotate faster (slower) than the spiral arms inside (outside) of the co-rotation resonance. 

Contrary to the expectations of much analytical work described above, hydrodynamic and gravity-only $N$-body simulations revealed a different scenario for spiral arms: transient, recurring spiral arms \citep[e.g.][]{Grand2012, GKC12,RFV13, Baba2013, Sellwood2014}. In this picture, individual spiral arms disappear on dynamical timescales while new spiral arms continually form, which ensures that the spiral morphology is long-lived. For example, \citet{Grand2012} and \citet{WBS11} found that spiral arms in their idealised $N$-body simulations are highly transient features that rotate at the same speed as stars everywhere: dubbed {\bf \emph{dynamic/co-rotating spiral arms}} (top-right panel of Fig.~\ref{fig:schem}). These spiral features appeared to grow via a mechanism akin to swing amplification \citep{Goldreich1965,JT66,T81}, in which the galactic shear rotation, orbit epicycle frequency of stars, and self-gravity conspire to amplify locally over-dense patches of material. Individual co-rotating spirals sweep out a range of pitch angles as they grow and decay on $\sim 100$ Myr timescales \citep[e.g.][]{WBS11,Baba2013,Grand2013}. Interestingly, the pitch angle attained at maximum spiral amplitude correlates with the rotation curve profile (or shear rate) such that lower (higher) shear rates of rising (decreasing) rotation curves exhibit more loosely (tightly) wound spiral arms. This matches the trend and scatter of the observed pitch angle-shear rate relation \citep{Grand2013,Michikoshi2016,Masters2019}. 

Another interpretation of the transient, recurring spiral arms seen in numerical simulations is {\bf \emph{multiple mode theory}}, which describes spiral arms as the superposition of multiple wave patterns with different discrete pattern speeds \citep{Minchev2012,Comparetta2012, Roskar2012,Sellwood2014,Kwak2025}. The top-middle panel of Fig.~\ref{fig:schem} shows the pattern speed profile expected for multiple mode theory: each spiral mode is characterised by a single pattern speed with power that lies within that pattern's inner and outer Lindblad resonance radii. Modes that dominate in the inner disc are faster compared to those that dominate the outer disc. Consequently, spiral modes can overlap in radius, leading to the constructive interference of modes as they pass one another to give the appearance of larger scale, winding spiral arms that grow and decay periodically on dynamical timescales \citep[e.g.][]{Comparetta2012,Minchev2012,Sellwood2014} - a phenomenon that some simulations suggest may cause non-linear growth of spiral over-densities \citep{Kumamoto2016}. Unlike the classic density wave, each of these modes need not be long-lived, but rather transient and recurring on $\sim$ Gyr timescales.

Other proposed mechanisms that may be at play in barred galaxies include bar-driven density waves \citep{Salo2010}, coupled bar-spiral modes \citep[e.g.][]{Hilmi2020}, and Manifold spirals \citep{RG07,Athanassoula2012}. The bottom-middle panel of Fig.~\ref{fig:schem} depicts the pattern speed profile for the {\bf \emph{Manifold theory}} of spiral arms \citep{RG07,ARM09,Athanassoula2012}, which provides a description of spirals and rings found in barred galaxies. The basis of manifold theory is that stars on unstable \citep{Lyapunov1949} orbits escape from Lagrange points at the ends of the bar major axis and move along regions of space defined by ``manifolds''. Because these stars originate from the bar, the spiral pattern that they produce has the same constant pattern speed as the bar itself. Therefore, a defining signature of manifold theory is a fixed pattern speed that extends from the centre of the galaxy out to radii beyond the bar where spiral structure exists.

\subsection{Observationally inferred pattern speeds}

The variation in spiral arm pattern speed predicted by the theories described above offers an observational test that can in principal elucidate the nature of spirals \citep[see Section 4 of][for a thorough review]{Dobbs2014}.  However, pattern speeds are difficult to measure directly in observed galaxies. The well-known Radial Tremaine-Weinberg (RTW) method \citep{MRM06} uses spatial and kinematic information of gas in inclined disc galaxies to solve the continuity equation, allowing $\Omega _p$ to vary radially. Studies that apply this method typically find that pattern speeds decrease with radius \citep[e.g.][]{MRM08a,MRM08,SW12}. However, whether such decreases are continuous or characterised by a series of distinct separate patterns, each with a constant pattern speed, remains debated \citep[e.g.][]{Speights2025}.

Another commonly used approach to infer spiral pattern speeds is the measurement of spatial offsets of various tracers of star formation; tracers of earlier and later stages of the star formation process should be located progressively farther (in the azimuthal direction) from density wave spiral arms. This is because the constant pattern speed of QSSS predicts the formation of a spiral shock in the gaseous component as gas flows through the spiral potential minimum, leading to the compression required to trigger star formation \citep{Roberts1969}. Because material flows tangentially across spiral arms everywhere except the co-rotation radius, this leads to a radially-dependent azimuthal offset between dense molecular gas (CO) and young star formation tracers (e.g H$\alpha$, UV). In contrast, no shock or offsets are predicted at any radius for dynamic (co-rotating) spiral arms. Based on such a clearly distinguishing prediction, one may expect observations to swiftly resolve the issue. However, a consensus is still lacking: there exist many observations of star-forming tracer offsets in various galaxies that support one theory or another \citep[e.g.][]{Egusa2009,FR10,Ferreras2012}. Most recently, \cite{Querejeta2025} analysed the spatial offsets of star-forming tracers in 24 nearby spiral galaxies from the PHANGS survey \citep{Querejeta2021}, and found that a variety of spiral arm theories were supported: most were consistent with either overlapping spiral modes \citep[e.g.][]{Sellwood2014} or transient dynamic (co-rotating) spirals \citep[e.g.][]{Baba2013}, and a minority were found to be consistent with a single quasi-stationary density wave \citep[e.g.][]{LS64}. There is also a lack of consensus for the Milky Way: for example, \citet{Hou2015} find observational indications supporting density waves, whereas \citet{Hunt2018} and \citet{Funakoshi2024} find support for dynamic spiral arms. The collective picture gathered from idealised simulations and observational surveys therefore suggests that a single theory is insufficient to explain the complexity of spiral morphology in the local Universe. 

\subsection{High resolution cosmological simulations}

With respect to theory, while numerical simulations have provided valuable insights into the mechanisms of spiral arm formation and evolution, they often have highly idealised initial conditions that omit the cosmological galaxy formation context. A promising avenue for furthering our understanding is to investigate spiral arms in cosmological hydrodynamic simulations. Although such studies are relatively scarce, cosmological simulations are beginning to predict observable spiral signatures, such as systematic streaming motions associated with spiral arms that induce detectable azimuthal metallicity variations \citep{Grand2016,SanchezMenguiano2016,Orr2023}. They are also being used to study pattern speeds directly: the recent study of \citet{Quinn2025} investigates the FIRE cosmological simulations of Milky Way-mass galaxies \citep{Wetzel2023} and shows evidence for multiple spiral modes. In contrast, \citet{Ghosh2025} study the spiral arms at $z=0$ in Milky Way-mass galaxies from the TNG cosmological simulations \citep{Pillepich2024} and claim that they are long-lived density waves. Thus, the study of spirals in cosmological galaxy formation simulations is in its infancy, yet it has already produced fascinating results that may point to a dependence of spiral arms on baryonic physics models in addition to the multiple potential physical mechanisms and perturbation sources.

To further complicate matters, numerical resolution has been shown to impact spiral arms found in $N$-body simulations. \citet{Fujii2011} ran several simulations of pure stellar discs with varying particle number, and found that at least $\sim10^6$ particles were needed to ensure the spiral pattern survives longer than 10 Gyr; fewer particles resulted in the disappearance of the spiral pattern within a few gigayears. Later, \citet{DO12} showed that spiral arms seeded by Poisson noise cease to form spontaneously in stellar discs modelled with $10^8$ particles; spirals formed only in response to heavy particles included by hand. Interestingly, these perturbations formed spirals initially via a swing amplification-like mechanism that later appeared to be highly non-linear and self-sustaining, raising the question of the impact of numerical resolution on the nature and longevity of spiral arms \citep[although see][for another point of view]{Sellwood2014}.

In this paper, we aim to shed further light on the nature of spiral arms by studying spiral arms in the high resolution \auriga \superstars cosmological simulations that model the stellar disc with $\sim 10^8$ particles (\citealt{Grand2023,Pakmor2025b}, and Fragkoudi et al. in prep.). This number of particles minimises numerical noise/heating \citep{Fujii2011,DO12} such that high-fidelity stellar dynamics is combined with the full cosmological setting, capturing the detailed dynamical response of cosmologically-grown discs to a spectrum of physical perturbations, such as gas accretion, stellar feedback, and the gravitational influence of dark matter subhalos. We organise our paper as follows. In Section 2, we describe the \auriga \superstars cosmological simulations, including a summary of the \auriga galaxy formation model and the \superstars method. We provide also a description of the methodology used to compute the spectrograms. In Section 3, we present our results on the spiral pattern speed profiles of our \auriga \superstars halos and their evolution at late times, and link them to existing theories of spiral structure. In Section 4, we discuss the implications of our findings, and we present our conclusions in Section 5.

\section{The Auriga Superstars simulations}
\label{sec:method}

In this paper, we study the \auriga \superstars simulations - a suite of cosmological magnetohydrodynamical simulations of Milky Way-mass halos, with 800 $\rm M_{\odot}$ per star particle resolution. These simulations use the \auriga galaxy formation model \citep[described in][]{GGM17} together with the \superstars method \citep[described in][]{Pakmor2025b} to significantly enhance the stellar particle resolution. We describe both of these aspects below.

\subsection{The Auriga galaxy formation model}

The Auriga galaxy formation model simulates the formation and evolution of galaxies in a $\Lambda$CDM universe from very early cosmic times ($z=127$) to the present day. It is implemented in the moving-mesh codes \textsc{arepo} \citep{Sp10,PSB15, Weinberger2020} and \textsc{arepo2} (see Section~\ref{sec2:suite} for more details), which solve the equations of magnetohydrodynamics on a moving Voronoi mesh with a second-order finite-volume scheme. This is fully coupled to self-gravity with a combined tree and particle-mesh solver. The model includes primordial and metal line cooling; a uniform UV background with reionisation completed by $z=6$; an effective model for the interstellar medium and stochastic star formation \citep{SH03}; magnetic fields \citep{PMS14,PGG17,PGP18}; an effective model for stellar feedback-driven galactic winds, mass return from stars from stellar winds and supernovae; and a model for the seeding, growth, and energetic feedback from black holes \citep[see][for more details]{GGM17}.  

The \textsc{auriga} model has been used to run suites of cosmological zoom-in simulations of Milky Way-mass galaxies and dwarf galaxies, which are publicly available \citep[see][for details]{Grand2024}. The Milky Way-mass simulations in particular have been extensively used to study the properties, origins, and evolution of central discs \citep{GKB20,Fragkoudi+Grand+Pakmor+19,GWG16}, stellar halos \citep{MGG19}, bulges \citep{GMG19}, bars \citep{Fragkoudi2021,Fragkoudi2025}, satellites \citep{SGG17}, magnetic fields \citep{PGG17}, dark matter \citep{PFG19,GW21}, and the circumgalactic medium \citep{vendevoort2021}. 

Most \auriga results achieved so far come from the standard so-called ``level 4'' resolution \auriga simulations: $m_b\sim5\times10^4\ \rm M_{\odot}$, $m_{\rm DM}\sim4\times10^5\ \rm M_{\odot}$, where $m_b$ and $m_{\rm DM}$ are the mass of the baryonic cell/particle and dark matter particle, respectively. \citet{Pakmor2025a} showed that increasing the resolution across dark matter, gas and stars leads to a systematic trend in galaxy properties, such as their stellar masses and star formation histories. However, from their test suite comprised of the same set of initial conditions at standard resolution run multiple times with different random number seeds \citep[also known as the Butterfly effect, see e.g.][]{Genel2019}, \citet{Pakmor2025a} showed that the intrinsic variance in the \auriga model for many galaxy properties is $\lesssim10\%$. This lends confidence that the model produces robust predictions of galaxy formation physics.

\subsection{The \superstars method and features}

To achieve a high stellar resolution for our cosmological simulations, the results of \citet[][]{Pakmor2025a} motivate an alternative approach to the traditional resolution increase of all matter components. To this end, we developed a new approach, called \auriga \superstars \citep[see][for a complete description of the method]{Pakmor2025b}, which 
increases the resolution of the star and dark matter particles while maintaining the same gas cell mass. This approach minimises systematic trends that emerge from increases in gas resolution and preserves the small intrinsic variance of the \auriga model found by \citet[][]{Pakmor2025a};  good convergence is achieved for galaxy properties such as the star formation history and radial profiles of surface density for improvements in stellar mass resolution by factors of 8 and 64\footnote{In Appendix.~\ref{sec:appendix}, we present a resolution study of the spiral arms of halo 6 simulated at different stellar resolutions.}. Moreover, this approach significantly reduces the computational cost compared to increasing the gas resolution by the same factors. It is therefore ideal to run a suite of Milky Way-mass simulations with high stellar resolution for the study of galactic dynamics. Below, we provide a summary of the main aspects.  

\subsubsection{Enhanced collisionless particle resolution}

The principle feature of the \superstars method is the increased resolution of dark matter and star particles relative to the gas cells, which are kept to the fixed standard resolution of $m_{\rm gas}\sim5\times10^4\ \rm M_{\odot}$. First, the dark matter mass is increased by a factor 8 from its standard value of $m_{\rm DM}\sim4\times10^5\ \rm M_{\odot}$ to $m_{\rm DM}\sim5\times10^4\ \rm M_{\odot}$. This is done by beginning the simulation with initial conditions created at this higher dark matter resolution, then de-refining the Voronoi mesh once it is created from the dark matter-gas split pairing at the starting redshift. Second, the star particle resolution is increased by a factor of 64 relative to the gas cell resolution, i.e., to $m_{\rm star} \simeq 800\ \rm M_{\odot}$. This is done through the stochastic model for star formation: in \textsc{auriga}, a gas cell that satisfies the criteria for star formation is given a positive star formation rate and a non-zero probability to convert its mass into a star particle. In \textsc{Superstars}, gas cells selected to model a star formation event convert their mass into 64 star particles, each with mass $m_{\rm gas} / 64$, and each given an initial position identical to their parent gas cell. Their velocities are set to the velocity of the parent gas cell plus a random isotropic component. This random component is sampled from a Gaussian centered on the velocity of their parent gas cell with a width set by either the sound speed of the parent gas cell or the local velocity dispersion of the cell and its direct neighbours, whichever is smaller. After sampling from this Gaussian, we subtract the mass-weighted velocity vector of the newborn star particles to ensure their total momentum is equal to that of their parent gas cell. These sibling star particles are unbound from one another -- as are open clusters -- and drift away from each other over time. The chemical enrichment, evolution, and stellar feedback is handled in exactly the same manner as the original \auriga simulations. 

It is important to note that \citet{Pakmor2025b} showed that the dark matter to stellar particle mass ratio, $m_{\rm DM}/m_{\rm {star}}=8$, in \auriga \superstars does not lead to systematic changes induced by artificial scattering of particles of the type shown by \citet{Ludlow2019,Ludlow2021}. This is likely because the relevant mass segregation timescale scales with the total number of particles, $N$, as well as $m_{\rm stars}/m_{\rm DM}$: the former is much larger in \auriga \superstars compared to lower-resolution large cosmological volume simulations.

\subsubsection{Refinement of gas and stars in dwarf galaxies}

In addition to the above tailored resolution increases for star particles and dark matter, our simulations also include refinement of gas cells belonging to dark matter subhalos with mass $< 10^{10} \rm M_{\odot}$ $h^{-1}$. For these subhalos, we mark gas cells found within a half-mass radius to be refined by a factor of 8 in mass (2 in length). This ``dwarf refinement'' is motivated by both better chemical evolution at low $Z$ and the results of \citet{GMP21}, who found that high gas resolution facilitated the formation of dozens of ultra-faint dwarf satellite galaxies with stellar mass $\lesssim 10^5\ \rm M_{\odot}$ which do not form at lower gas resolution. 

In order to preserve the stellar particle mass of $800\ \rm M_{\odot}$ across the simulation, only 8 star particles are spawned from a ``dwarf-refined'' star-forming gas cell instead of the usual 64 particles spawned in any other star-forming gas.

\subsection{The \auriga \superstars simulation suite}
\label{sec2:suite}

The \auriga \superstars simulation suite consists of 10 halos that were run as part of the \auriga project \citep[][]{GGM17}, and will be described in full in Fragkoudi et al. in preparation. The initial conditions were constructed via the method of \citet{J13} for relatively isolated Milky Way-mass objects selected from the EAGLE dark matter only simulation of comoving side length 100 cMpc \citep[L100N1504, introduced in][]{SCB15}. The adopted cosmological parameters are $\Omega _m = 0.307$, $\Omega _b = 0.048$, $\Omega _{\Lambda} = 0.693$, $\sigma_8 = 0.8288$, and a Hubble constant of $H_0 = 100\, h\, \rm km \, s^{-1} \, Mpc^{-1}$, where $h = 0.6777$, taken from \citet{PC13}. Detailed information about the initial conditions and the procedure used to generate them can be found in \citet{Grand2024}. As noted in the previous section, the \auriga \superstars initial conditions have a dark matter particle resolution of $5\times10^4\ \rm M_{\odot}$, which is equivalent to that of the high resolution \texttt{Original}/3 simulation set described in \citet{Grand2024}. In this paper, we present results from three of the \auriga \superstars halos which correspond to the objects labelled halos 6, 18, and 25 of the original \auriga simulation suite. 

The \auriga \superstars simulations are run with all the features described above implemented into \textsc{arepo 2}, which includes several additional advantages compared to the original \textsc{arepo} code. Among these are: a faster and more accurate gravity solver; SUBFIND-HBT, which is a faster and more accurate halo finder; an on-the-fly merger tree that eliminates the need to post-process merger trees; and an on-the-fly movie generator. 

In addition to the 256 full snapshots and group catalogues, our \auriga \superstars simulations include in their output the positions and velocities of star particles at a high-cadence output rate of 5 Myr. This high temporal resolution is ideal for tracking stellar orbits and analysing the evolution of galactic stellar structures. In this paper, we will make use of this output by performing an analysis of the formation and evolution of spiral arms, in particular their pattern speeds.

\subsection{Fourier analysis of spiral structure}
\label{sec:fourier}

We calculate spectrograms from a Fourier analysis of spiral structure in order to analyse the radial profiles of spiral pattern speeds. To begin, we define the coordinate system such that the particle data is centred on the main galaxy centre for each high-cadence snapshot. To do this, we interpolate the main galaxy centre coordinates from the time series of galaxy coordinates given by SUBFIND-HBT for the 256 full snapshots. We then use these interpolated coordinates as the starting point for a ``skrinking spheres'' approach applied to star particles to refine the coordinates further: we find that 4 iterations is sufficient for convergence. To rotate the galaxy, we first calculate the angular momentum vector from all star particles inside 20 kpc, and normalise it with respect to the total angular momentum to give the angular momentum unit vector, $\hat{L}\equiv \hat{i}$. We then define the unit vectors $\hat{j}=\hat{i}\times \hat{a}$ and $\hat{k}=\hat{i}\times\hat{j}$, where $\hat{a}=(1, 0, 0)$. The unit vectors $\hat{i}$, $\hat{j}$, and $\hat{k}$ comprise a rotation matrix which we apply to all particle vector coordinates, including particle positions and velocities. Finally, we transform from cartesian to polar cyldinrical coordinates.

The strength of an $m-$symmetric spiral structure at a radius, $R$, can be quantified using the coefficient:

\begin{equation}
    c_m (R) = \frac{1}{2\pi} \int _{-\pi}^{\pi} M (R, \theta)  e^{im \theta} d\theta,
\end{equation}
where $M(R,\theta)$ is the mass at the coordinate $(R,\theta)$. Because we are working with discrete particle data specified at discrete times,  we select all star particles with a vertical height less than 1 kpc from the galactic mid-plane and within a cylindrical radius of 25 kpc for each snapshot. We bin these selected particles in 1 kpc-wide radial annuli, and calculate

\begin{equation}
    c_m (R) = \sum _i^N m_i \bigg[\cos{(m\theta _i)} + i\sin{(m\theta _i)}\bigg],
\end{equation}
for each annulus. Here, $N$ is the total number of star particles in the annulus, and $m_i$ and $\theta _i$ are the mass and azimuthal coordinate of the $i$-th star particle, respectively. This quantity is often used to define the normalised amplitude, $A_m = |c_m|/|c_0|$ \citep[see, e.g.][and the resolution study presented in appendix.~\ref{sec:appendix}]{Grand2013}.

Under the assumption that a spiral arm can be represented as a wave mode(s) that rotates from time, $t_0$, to some later time, $t_1$, we define

\begin{equation}
    \tilde{c}_m (R,\omega) = \int _{t_0}^{t_1} c_m(R,t) e^{i\omega t} dt,
\end{equation}
where $\omega=m\Omega _p$ is the rotation frequency linked to the pattern speed, $\Omega _p$, during this time period. 

In lieu of calculating $\tilde{c}_m$, we calculate the discrete Fourier transform, 

\begin{equation}
    C_{k,m}(R) = \sum _{j=0}^{N-1} c_m (R,t_j) w_j e^{2\pi ijk/N}; k=0, ..., N-1 ,
\end{equation}
where $N$ is the number of data points comprising the signal, $k$ represents a discretised frequency bin, and $w_j$ is the Hann window function:

\begin{equation}
    w_j = \frac{1}{2} \biggl[ 1 - \cos (2\pi j / N) \biggr].
\end{equation}
The purpose of a Window function of this kind is to minimise spectral leakage, which can occur if the underlying frequency of the data signal lies between two discretely sampled frequencies. This is because discretely sampled data points spanning a finite period of time is effectively an infinite signal multiplied by a square window function. The Fourier transform of this top-hat function is a sinc function with lobes of substantial spectral leakage into frequencies far from the true value. The use of a Hann-like window function smoothly tapers the input signal such that it gradually goes to zero outside of the sampling range, which significantly reduces and confines spectral leakage around the true frequency \citep{Press2002}. 

\begin{figure*}
\centering
\includegraphics[scale=0.5,trim={1.cm 7cm 2.5cm 3cm}, clip]{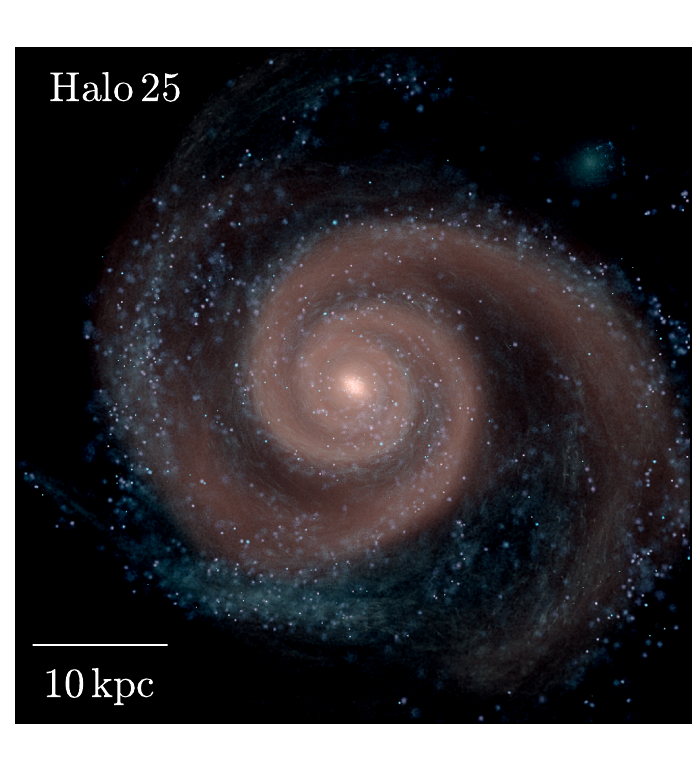}
\includegraphics[scale=0.5,trim={1.cm 7cm 2.5cm 3cm}, clip]{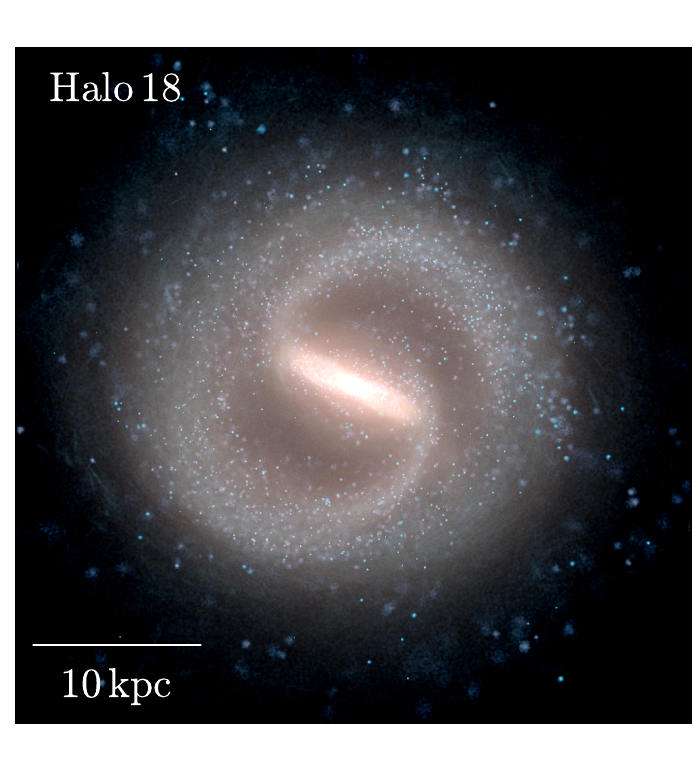}
\includegraphics[scale=0.5,trim={1cm 7cm 2.5cm 3cm}, clip]{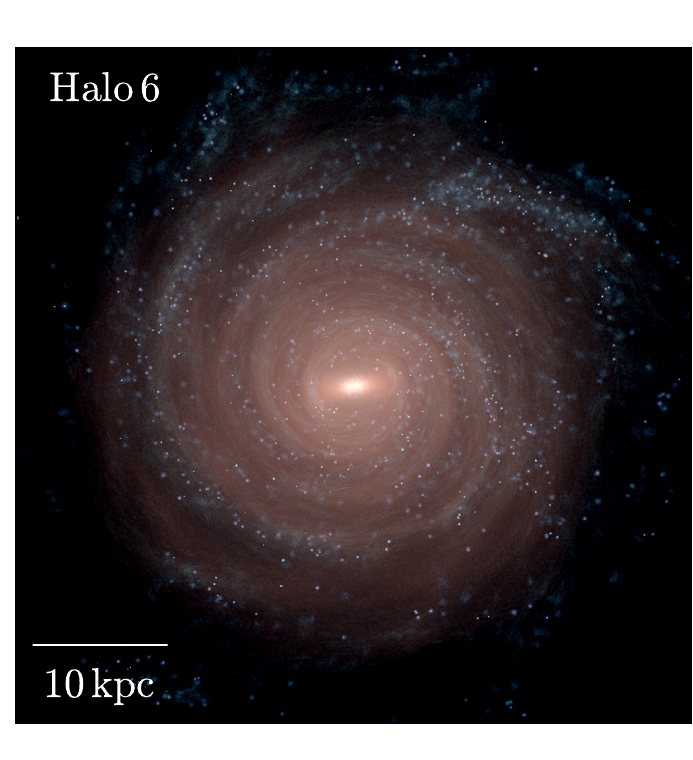}
    \caption{Face-on projected stellar light images for three \auriga \superstars simulations each illustrating a qualitatively different type of spiral structure (see text for details): halo 25 (left panel; $z=0$); halo 18 (middle panel; $z\sim 0.25$); and halo 6 (right panel; $z=0$). The images are synthesized from a projection of the K-, B- and U-band luminosity of stars, which are shown by the red, green and blue colour channels, in logarithmic intervals, respectively. Younger (older) star particles appear bluer (redder).}
    \label{fig:lightproj}
\end{figure*}

We calculate the power in each frequency as:

\begin{equation} \label{eq1}
\begin{split}
P_m(f_0) & = \frac{1}{W^2} |C_{0,m}|^2 \\
P_m(f_k) & = \frac{1}{W^2} \bigl[|C_{k,m}|^2 + |C_{N-k,m}|^2 \bigr] \\
P_m(f_c) & = \frac{1}{W^2} |C_{N/2,m}|^2
\end{split}
\end{equation}
where in the middle line $k=1, 2, ..., (N/2 - 1)$, 

\begin{equation}
    W = N \sum _{j=0}^{N-1} w_j^2,
\end{equation}
is a normalisation factor to take into account the window function, and 

\begin{equation}
    f_k = 2f_c \frac{k}{N}
\end{equation}
is the $k$-th disretely sampled positive frequency: $k=0, 1, ..., N/2$. Here, $f_c = (N-1) / (2T)$ is the Nyquist frequency defined by the number of samples, $N$, and the total length of the time window, $T$.

In this study, we perform this analysis on time series of $N=256$ snapshots separated by $\Delta =5$ Myr each. This gives a Nyquist frequency of $f_c=100 \pi$ $\rm rad \, Gyr^{-1}$ $\approx 100 \pi$ $\rm km \, s^{-1} \, kpc^{-1}$ for an $m=2$ pattern. Given that no spiral pattern is able to rotate as quickly as this, our sampled data should contain complete spectral information and no aliasing of power from frequencies larger than the Nyquist frequency should be present in our spectrograms. The time window of our spectrograms is $T=N\Delta=1.28$ Gyr, which gives a frequency resolution of $2\pi/(1.28m)$ $\rm rad \, Gyr^{-1}$, which is $\sim 2.4$ $\rm km \, s^{-1} \, kpc^{-1}$  for an $m=2$ pattern.

\section{Results}

\begin{figure*}
\centering
\includegraphics[scale=0.5,trim={0.cm 0 1.5cm 0}, clip]{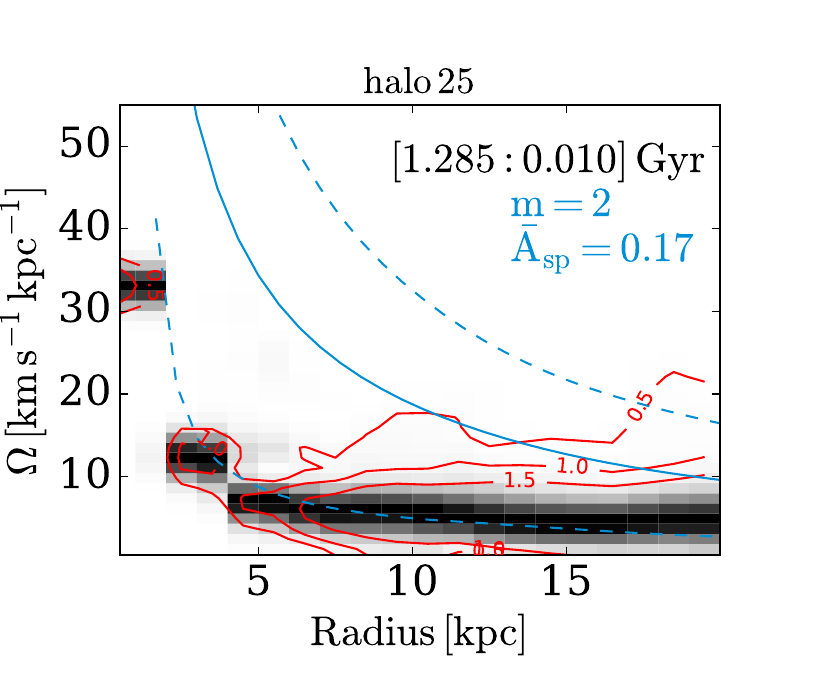}
\includegraphics[scale=0.5,trim={1cm 0 1.5cm 0}, clip]{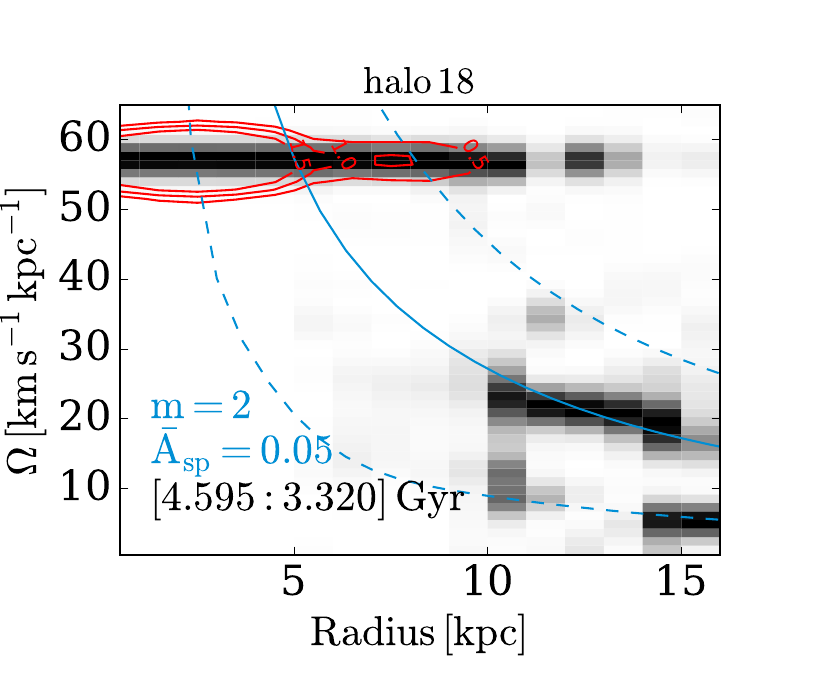}
\includegraphics[scale=0.5,trim={1cm 0 1.5cm 0}, clip]{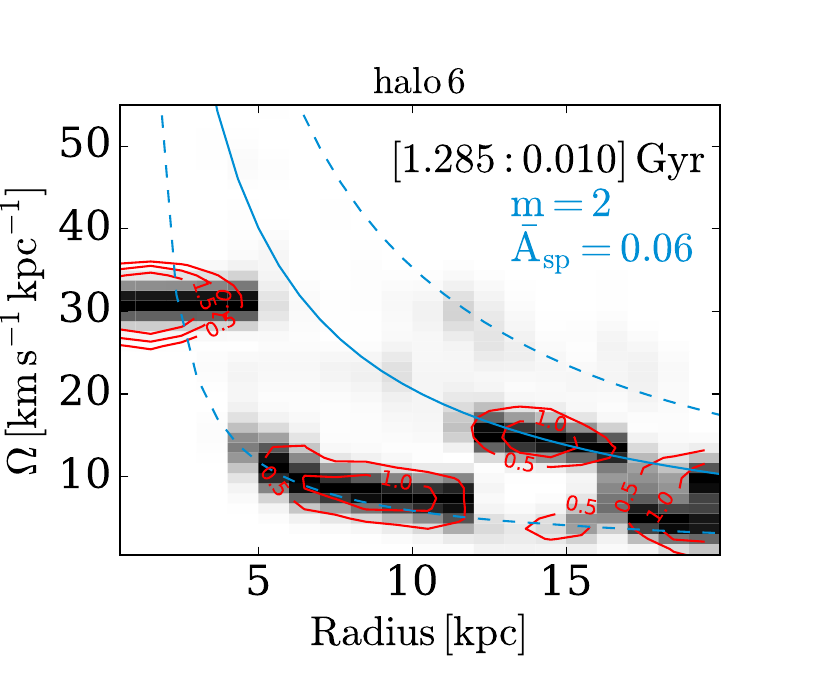}\\
\includegraphics[scale=0.46,trim={0.cm 0.5cm 2.1cm 0}, clip]{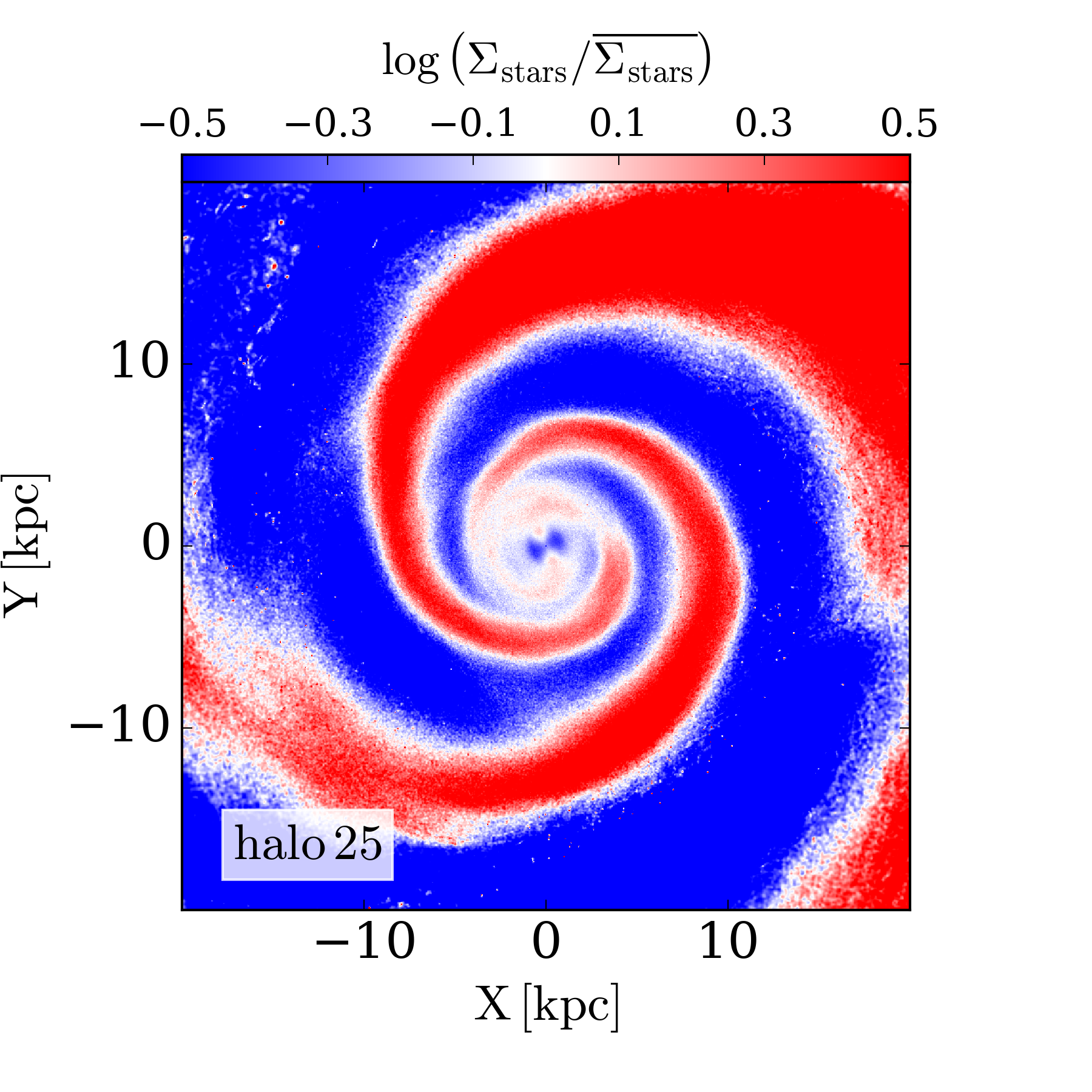}
\includegraphics[scale=0.46,trim={1.cm 0.5cm 2.1cm 0}, clip]{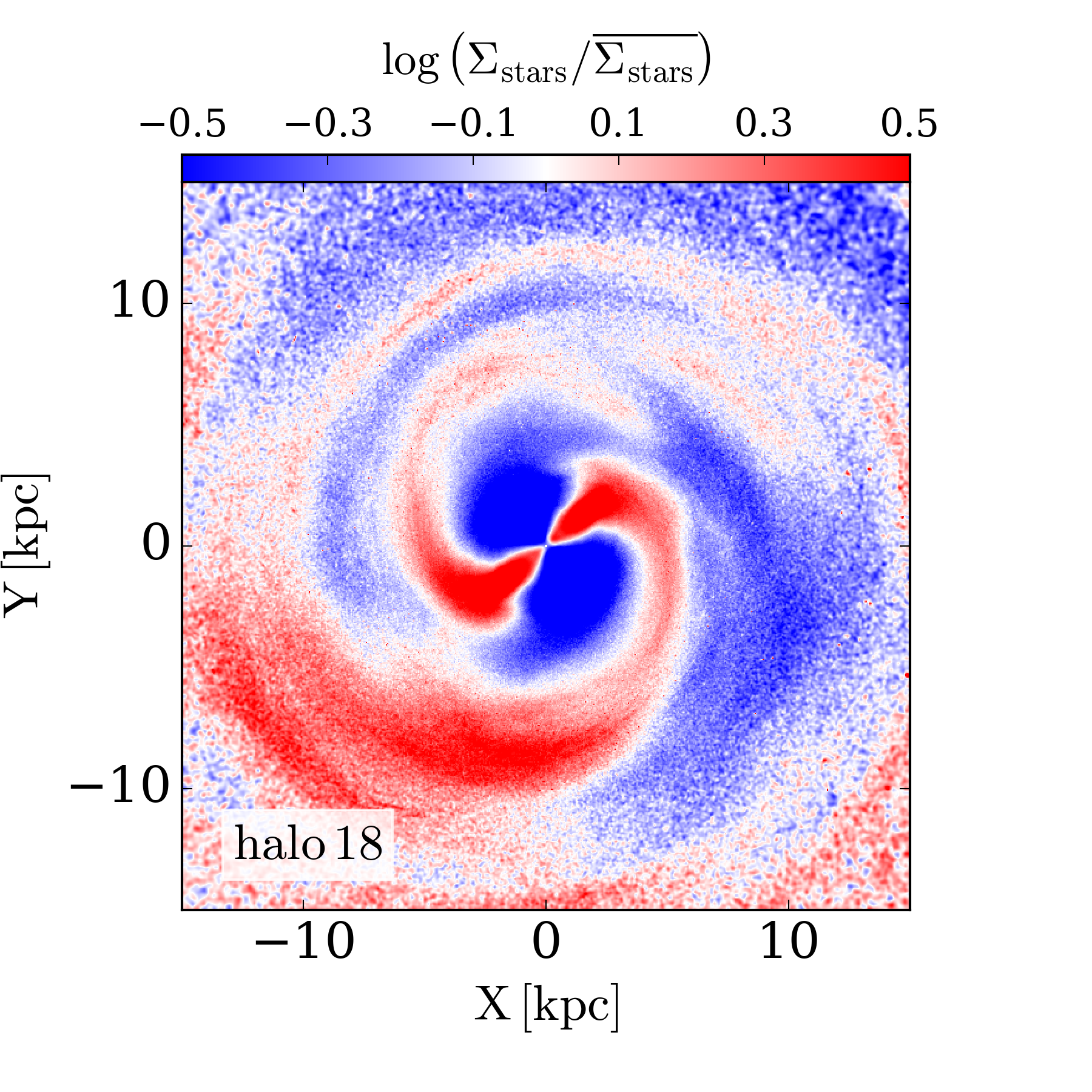}
\includegraphics[scale=0.46,trim={1.cm 0.5cm 2.1cm 0}, clip]{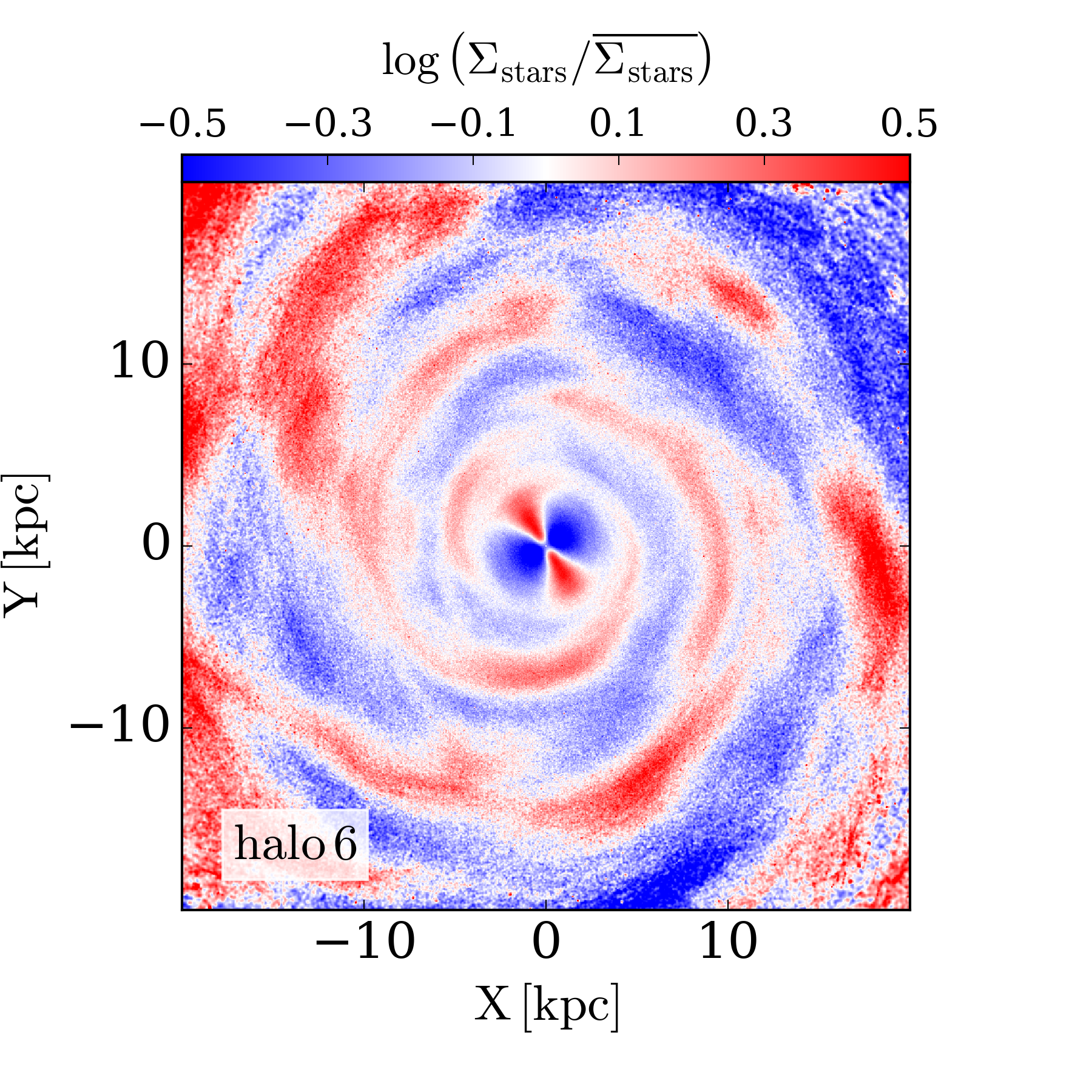}
\caption{{\it Top row}: Spectrograms of the $m=2$ Fourier mode for a galaxy with a strong tidal interaction (halo 25; left), a strongly-barred galaxy (halo 18; middle), and a weakly-barred galaxy (halo 6; right) at late times (as indicated in each panel). The greyscale histogram shows the column-normalised power; the power in each frequency and radial bin is divided by the maximum power found in the corresponding radius. Contours show the power distribution with a common (fixed) normalisation. The average amplitude for the spiral structure, $A_{\rm sp}$, for the corresponding time window is given in each panel. The angular rotation curve measured at the middle of the time window is plotted as solid curves; the inner and outer Lindblad resonance loci are plotted as dashed curves. {\it Bottom row}: For each \auriga \superstars halo, the face-on stellar azimuthal over-density projections at times near the middle of the corresponding spectrogram time window shown above. The colour bar indicates the logarithmic over-density. The galaxy in halo 25 (left column) experiences a strong tidal interaction with a massive flyby. This generates a powerful, large-scale grand-design $m=2$ spiral structure whose pattern speed is very low and closely follows the ILR, which aligns with the rotation profile of kinematic density waves. The galaxy in halo 18 (middle column) shows a clear and strong bar over-density in the central regions, with 2-armed spirals emanating from their ends. This dominant $m=2$ mode shows a single pattern speed of $\sim 55$ $\rm km \, s^{-1} \, kpc$ that stretches from the centre of the galaxy to (even slightly beyond) the OLR; the spirals have the same pattern speed as the bar, as expected for manifold theory. The weakly-barred disc galaxy in halo 6 (right) is relatively undisturbed at late times, and exhibits a coherent but complex spiral structure whose pattern speed profile appears to show a mixture of kinematic density waves concentrated mainly in the inner/middle disc and a dynamic spiral in the outer disc. These three clear examples demonstrate the wide-range of spiral structure types found in our simulation suite. }
    \label{fig:spectrogramall}
\end{figure*}

Fig.~\ref{fig:lightproj} shows the face-on stellar light projections for three \auriga \superstars simulations on which we focus in this paper. The images are very similar to the equivalent projections of the original Auriga counterparts shown in \citet{GGM17}. However, the superior mass resolution of \auriga \superstars enables finer sampling of the phase space {\it and gravitational potential}, which leads to sharper spiral/bar structures in comparison to their original counterparts (see Appendix.~\ref{sec:appendix} for a resolution study of the non-axisymmetric structure). We refer the reader to \citet{Pakmor2025b} for a discussion of these effects and how they affect accreted stellar debris structures as well as discs.

The three simulated galaxies presented in Fig.~\ref{fig:lightproj} are chosen because: i) they demonstrate the range of spiral/bar morphology within the suite; and ii) they each demonstrate a qualitatively different type of spiral arms (as we will show in Sec.~\ref{sec:3.2}). Firstly, halo 25 (left panel of Fig.~\ref{fig:lightproj}) shows a pronounced $m=2$ quasi-symmetric strong spiral arms reminiscent of ``grand-design'' spirals -- these arise in response to a strong tidal interaction with a satellite flyby \citep[see e.g.][]{Gomez2021}. Secondly, halo 18 (middle panel of Fig.~\ref{fig:lightproj}) has a strongly-barred disc with tightly wound spiral structure emanating from the ends of the bar major axis. Thirdly, the spiral arms of the weakly-barred disc of halo 6 (right panel of Fig.~\ref{fig:lightproj}) are more flocculent and not continuously connected across the full range of radii - this is the most common disc morphology found within the \auriga \superstars simulation suite. In what follows, we perform a quantitative analysis of the evolution of the spiral arms in these three simulated galaxies.

\subsection{A variety of spiral arm types in \auriga \superstars}
\label{sec:3.2}

In this section, we examine the spiral arms in the \auriga \superstars simulations and compare the radial profiles of their pattern speeds to those associated with different theories. We apply the Fourier analysis technique described in Section~\ref{sec:fourier} to a series of time windows which we initially centre on $\sim 4$ Gyr lookback time, and slide forward in time until the end of the last window is equal to $z=0$. In what follows, we present the results of selected spectrograms to highlight our main findings.

The top row of Fig.~\ref{fig:spectrogramall} shows spectrograms of the dominant $m=2$ Fourier component at late times for three \auriga \superstars simulations, each of which exhibits qualitatively different spiral pattern speed profiles as a function of radius. To aid in the interpretation of these spectrograms, the lower row of Fig.~\ref{fig:spectrogramall} shows the face-on map of the logarithmic azimuthal stellar over-density for a snapshot during the respective spectrogram time window in each case, in order to highlight density enhancements corresponding to bars and spiral arms\footnote{This is computed by calculating the azimuthally averaged stellar surface density in concentric radial annuli, which we use to normalise the surface density at each coordinate at the corresponding radius.}.

The top-left panel of Fig.~\ref{fig:spectrogramall} shows the spectrogram for the galaxy with the strong tidal interaction (halo 25). The power appears to cluster around the ILR curve at almost every radius, which indicates that the spiral arms in this galaxy are kinematic density waves. The bottom-left panel clearly shows a very strong $m=2$ grand-design morphology representative of the spiral structure within the last Gigayear of evolution. The onset of this coherent pattern is correlated with the flyby of a massive satellite, which produces also a global corrugation pattern in the disc \citep{Gomez2021}.

The upper-middle panel of Fig.~\ref{fig:spectrogramall} shows the spectrogram of the $m=2$ Fourier component for the strongly barred galaxy halo 18: the Fourier amplitude of the bar at this time is $\sim 0.5$. This spectrogram shows clear and coherent power for a pattern speed of $\Omega _p \sim 55$ $\rm km \, s^{-1} \, kpc$ from the galactic centre out to (even slightly beyond) the OLR at around 8--10 kpc radius. The bar length is close to the radius of co-rotation ($\sim 5$ kpc), which is consistent with the simulation of this halo at standard Auriga resolution \citep{Fragkoudi2021,Merrow2024}. The lower-middle panel shows the stellar over-density at an instance during this time window and reveals a clear quadrupole structure with a logarithmic overdensity larger than 0.5, which represents one of the strongest bars in the \auriga \superstars simulations. Spirals appear to emanate from the ends of the bar and form a quasi-ring like shape -- that is very reminiscent of manifolds such as those shown by \citet{RG07} -- contained within the radius of $\sim10$ kpc (outside of this radius, a separate, much weaker pattern speed profile following the rotation curve is found). Therefore, the pattern speed of the bar found inside co-rotation is the same as the pattern speed of the spiral arms outside of co-rotation. The pattern speed profile of this strongly barred galaxy during this epoch is thus consistent with the manifold theory for spiral arms, in which spirals are manifestations of stars flowing along manifolds after they leave an unstable Lagrange point at the end of the bar. 

The right column of Fig.~\ref{fig:spectrogramall} shows an example of the weakly barred galaxy, halo 6 (the Fourier amplitude of the bar at this time is $\sim 0.3$\footnote{Following \citet{Fragkoudi2025}, we consider the Fourier amplitude threshold qualifying a disc to be characterised as ``barred'' to be 0.25. As the Fourier amplitude of the bar in halo 6 lies just above this threshold, it is described as weakly barred.}) during an undisturbed period of evolution. The lower-right panel shows a clear but broken spiral structure extending out to radii of $\sim20$ kpc. The upper-right panel shows the $m=2$ spectrogram, which exhibits a more complex structure than those of halos 25 and 18: power is clustered along the ILR for the radial range $\sim 5$-$12$ kpc and $\sim 17$-$20$ kpc (indicating kinematic density waves at these radii); however, for radii $\sim 12$-$17$ kpc, the power clusters along the circular rotation curve. This spectrogram therefore suggests that kinematic density waves and dynamic spirals exist contemporaneously during this time window, wherein the latter overtake the former. This behaviour could explain the broken spiral appearance in the lower-right panel of Fig.~\ref{fig:spectrogramall}. We will analyse the evolution of spiral arms in this halo in Section.~\ref{sec:evolution} below.

In summary, Fig.~\ref{fig:spectrogramall} shows that there is a range of spiral arm types present in the \auriga \superstars simulations at late times: our chosen examples\footnote{The examples shown are broadly representative of the spiral arm types in the suite; most simulations are of the more flocculent type highlighted in halo 6.} show that a strong tidal interaction and a strong bar lead to pattern speed profiles consistent with kinematic density waves and manifold spirals, respectively. Our third example has neither a strong tidal interaction nor a strong bar, and exhibits a complex spiral pattern speed profile that may mean that more than one type of spiral structure is present in the time window in question. We also note that the amplitude of the spirals in the strong tidal interaction case (halo 25) is far stronger than the amplitude of spirals in the other cases (this is particularly evident in the lower panels of Fig.~\ref{fig:spectrogramall}).

\subsection{Evolution of spiral arms in \auriga \superstars}
\label{sec:evolution}

\begin{figure*}
\centering

\includegraphics[scale=1.,trim={1cm 0 0 1.5cm}, clip]{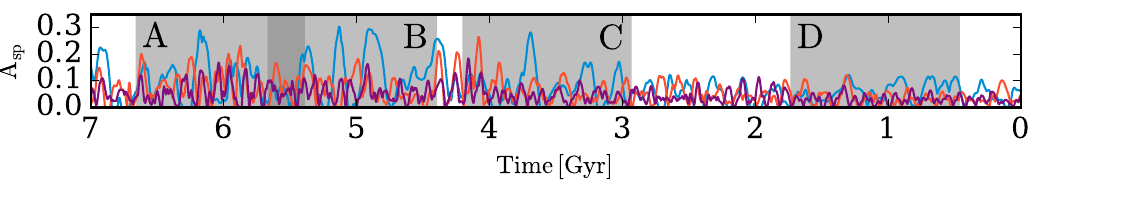}\\

\includegraphics[scale=0.5,trim={0 2cm 1.5cm 1.7cm}, clip]{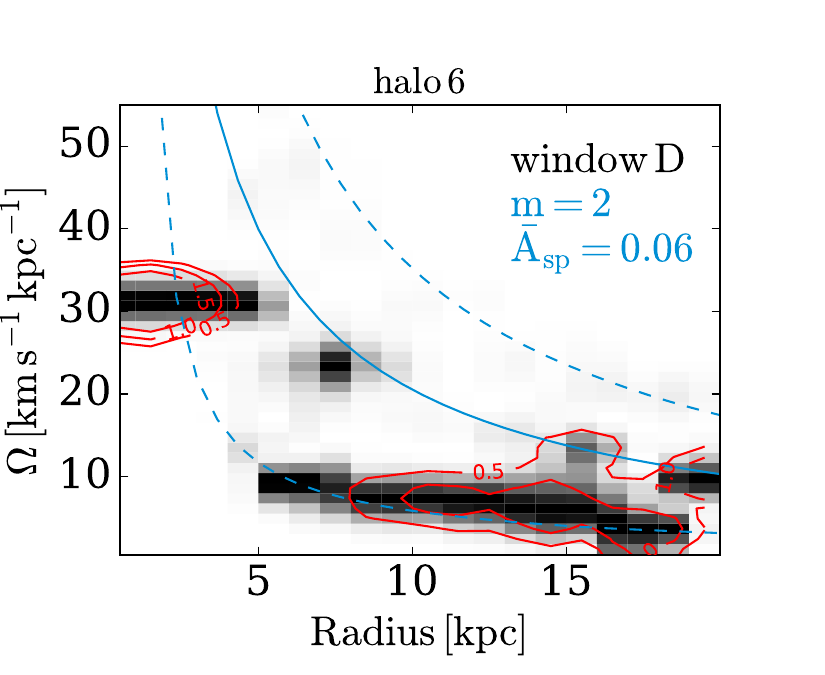}
\includegraphics[scale=0.5,trim={2cm 2cm 1.5cm 1.7cm}, clip]{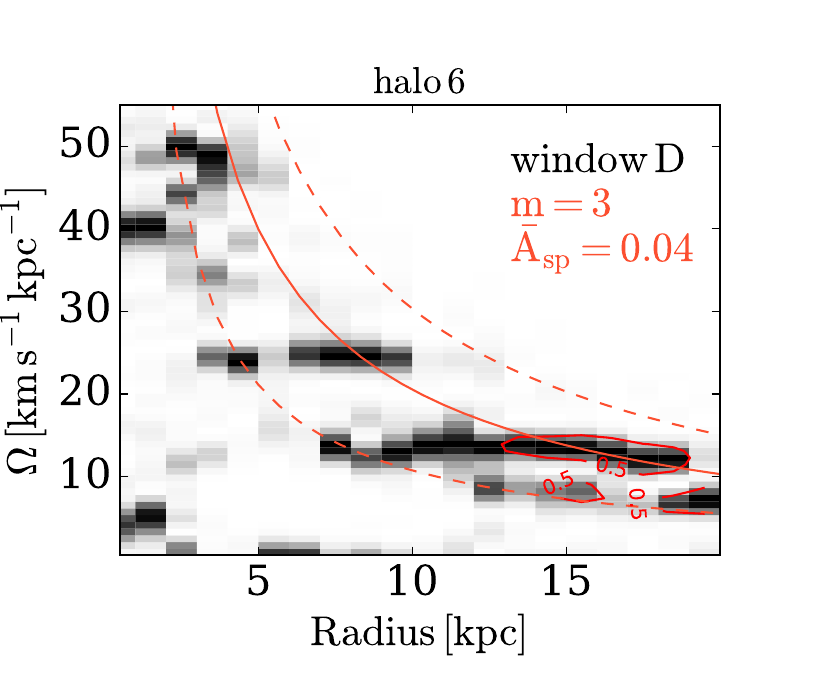}
\includegraphics[scale=0.5,trim={2cm 2cm 1.5cm 1.7cm}, clip]{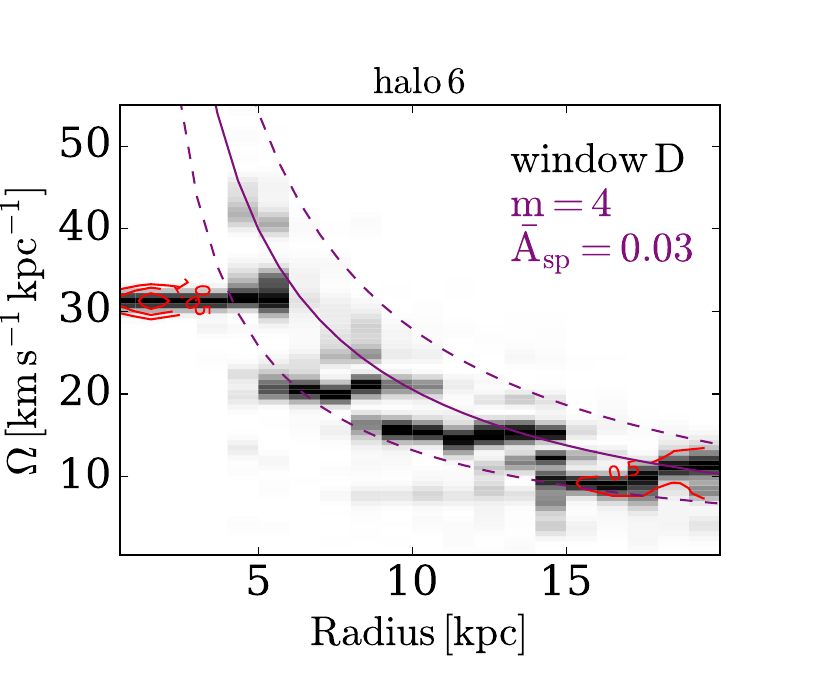}\\

\includegraphics[scale=0.5,trim={0 2cm 1.5cm 1.7cm}, clip]{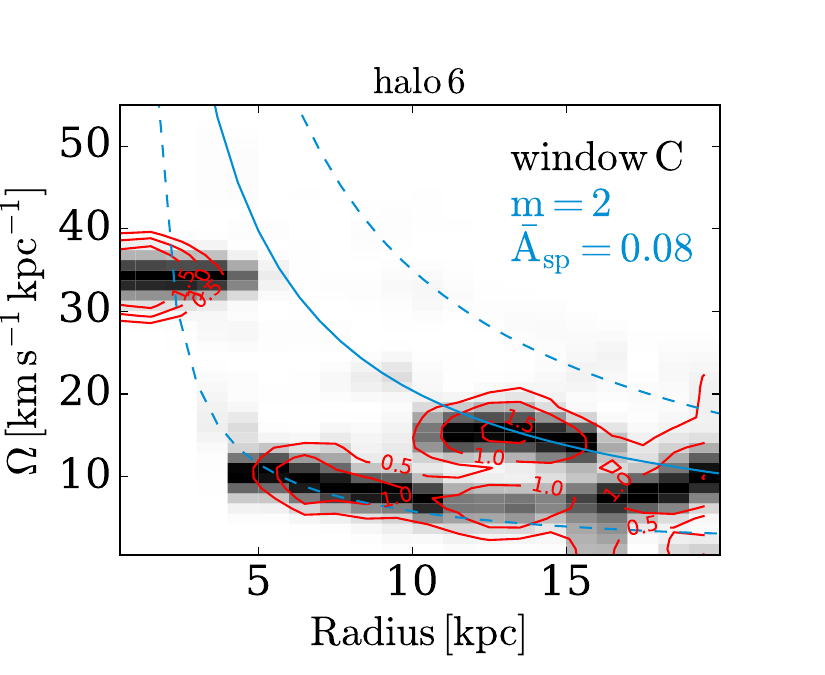}
\includegraphics[scale=0.5,trim={2cm 2cm 1.5cm 1.7cm}, clip]{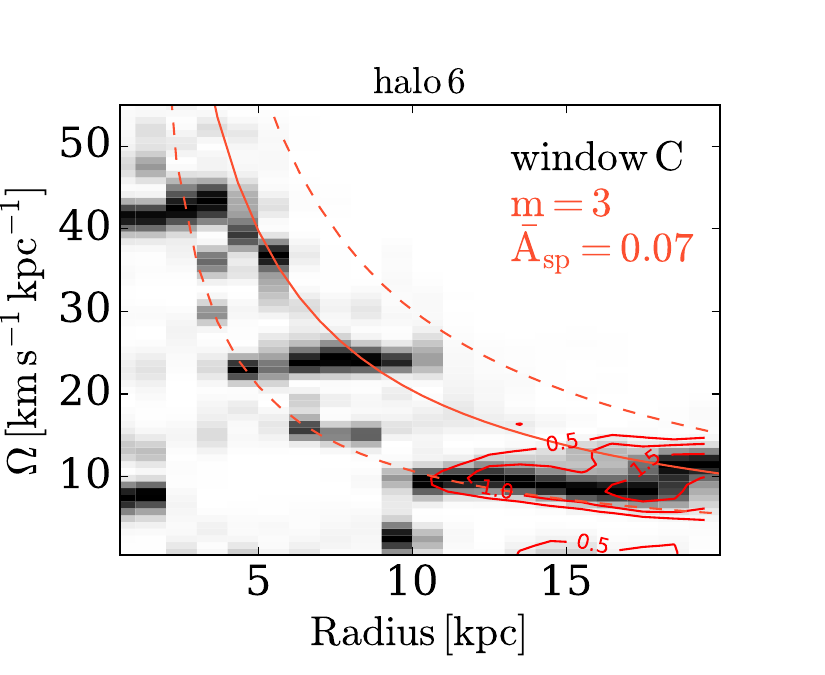}
\includegraphics[scale=0.5,trim={2cm 2cm 1.5cm 1.7cm}, clip]{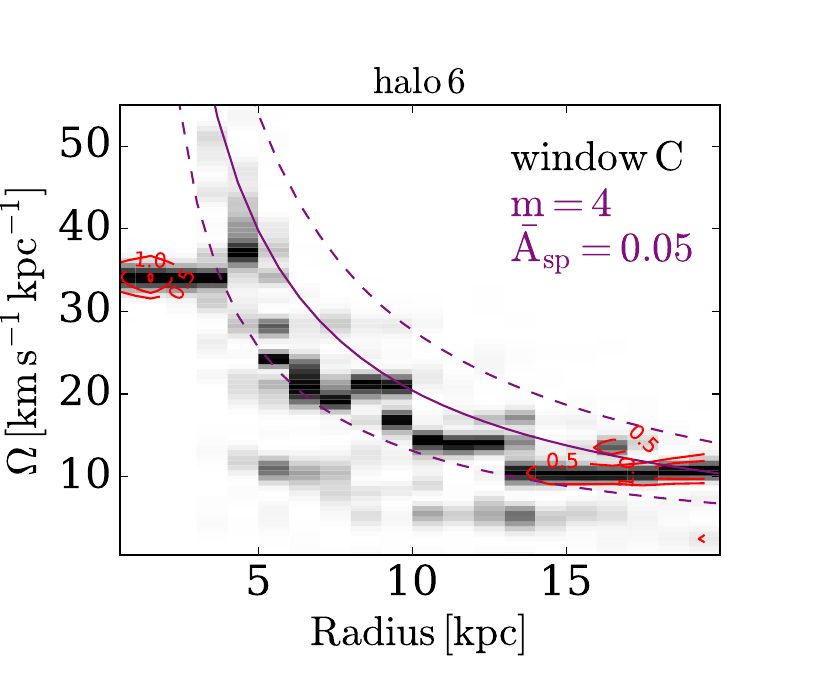}\\

\includegraphics[scale=0.5,trim={0 2cm 1.5cm 1.7cm}, clip]{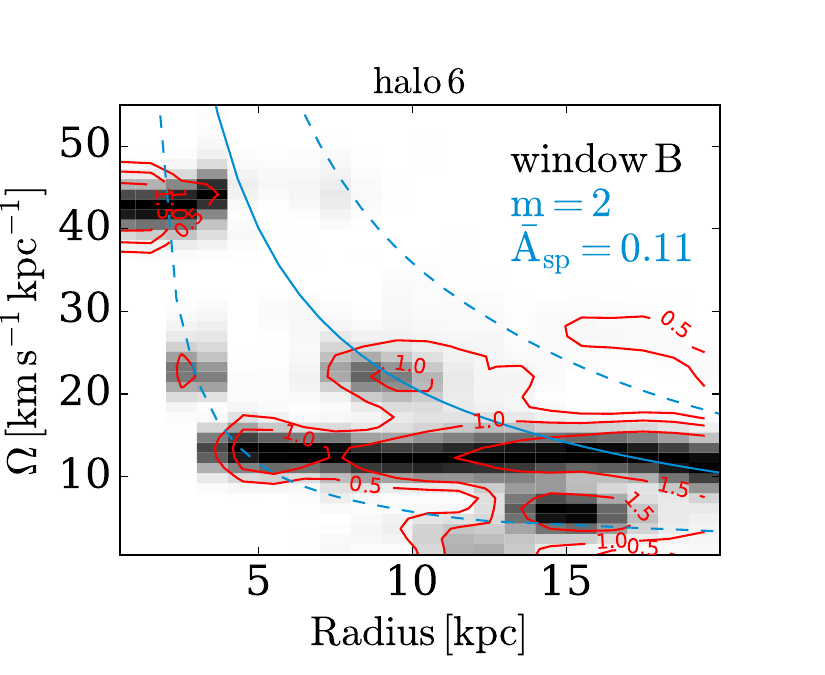}
\includegraphics[scale=0.5,trim={2cm 2cm 1.5cm 1.7cm}, clip]{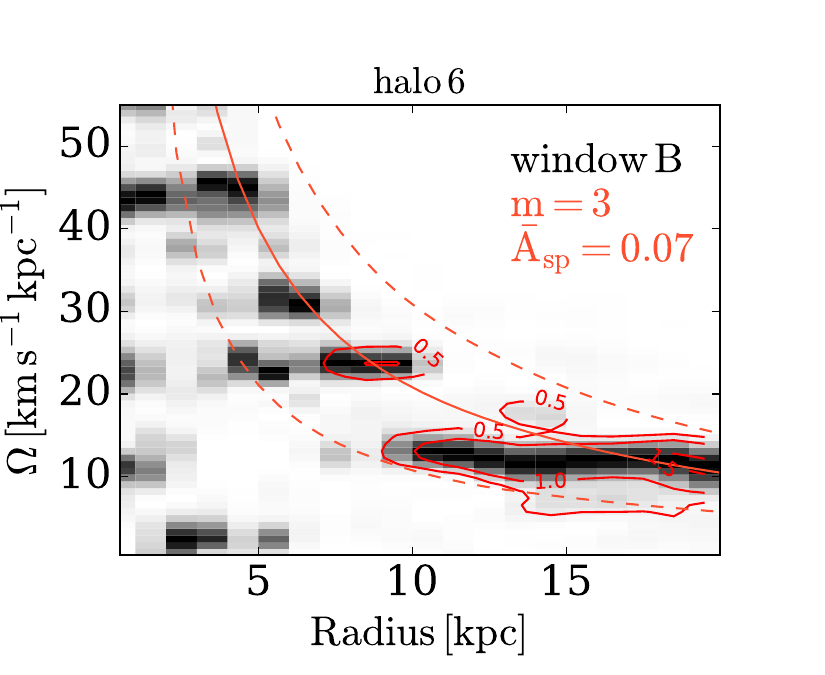}
\includegraphics[scale=0.5,trim={2cm 2cm 1.5cm 1.7cm}, clip]{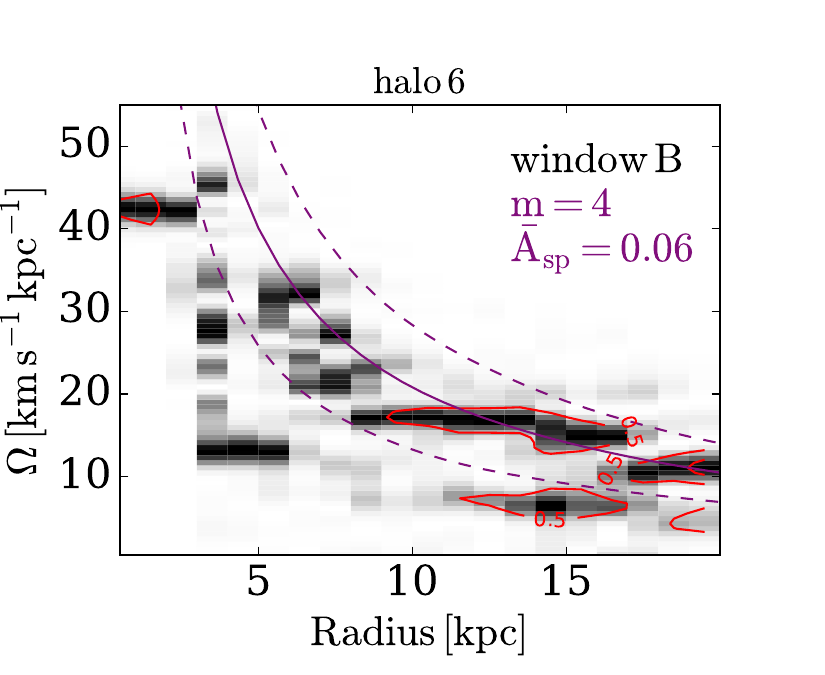}\\

\includegraphics[scale=0.5,trim={0 0 1.5cm 1.7cm}, clip]{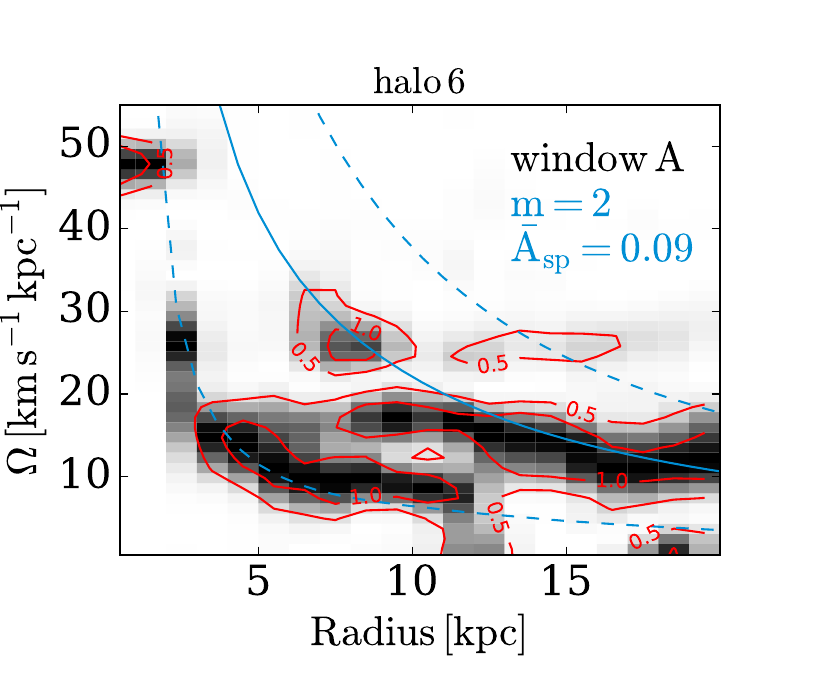}
\includegraphics[scale=0.5,trim={2cm 0 1.5cm 1.7cm}, clip]{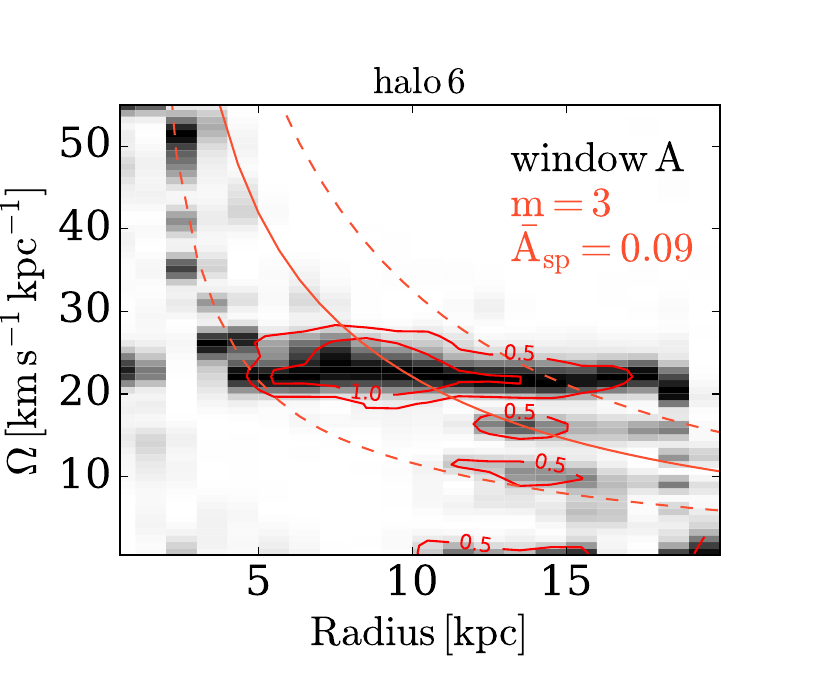}
\includegraphics[scale=0.5,trim={2cm 0 1.5cm 1.7cm}, clip]{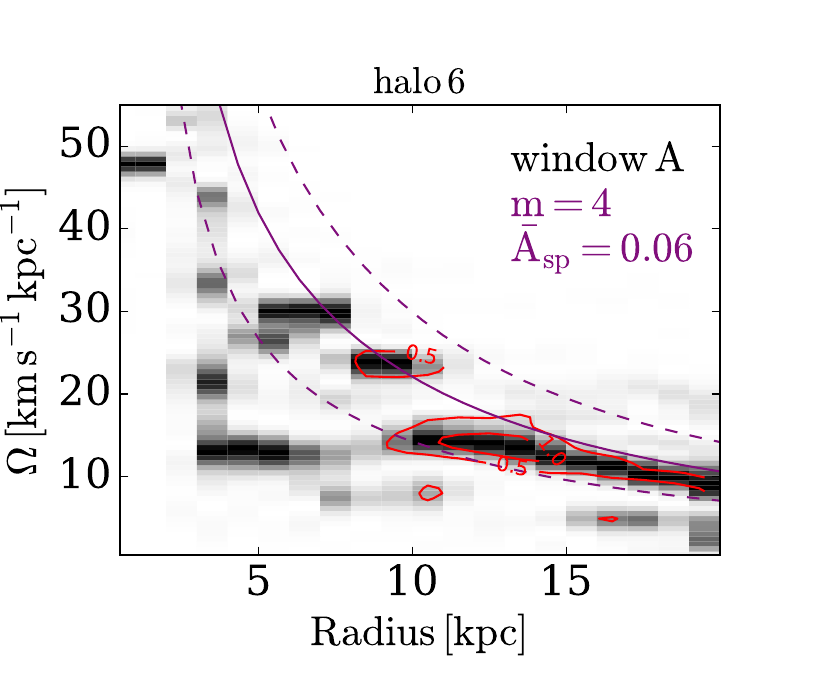}\\

\caption{Top panel: evolution of the amplitude of spiral modes within the radial range $7<R/{\rm kpc}<15$ for halo 6. Shaded regions indicate the time windows of the spectrograms shown in the lower panels, where the left, middle, and right columns correspond to the $m=2$, $m=3$, and $m=4$ Fourier modes, respectively. Different rows correspond to different time windows, as indicated at the top-right corner of each panel. The greyscale histogram shows the column-normalised power; the power in each frequency and radial bin is divided by the maximum power found in the corresponding radius. Contours show the power distribution with a common (fixed) normalisation. The angular rotation curve measured at the middle of the time window is plotted as solid curves; the inner and outer Lindblad resonance loci are plotted as dashed curves. Each panel indicates the time-averaged amplitude, $A_{\rm sp}$, for the corresponding time window and mode, $m$. This figure demonstrates the wide variety of behaviours that spiral arms evolve through for {\it the same galaxy}: i) in the 2nd row (window D), the dominant $m=2$ mode shows a kinematic density wave with power following the ILR at almost all radii; ii) in the 3rd and 5th rows (windows C and A), the $m=2$ spectrogram profile indicates an inner kinematic density wave and dynamic spirals at intermediate/outer radii; iii) in the 4th row (window B), $m=2$ mode shows a large-scale density wave with a constant pattern speed of $\sim 14$ $\rm km \, s^{-1} \, kpc^{-1}$; iv) the $m=3$ and $m=4$ modes sometimes show evidence of sub-dominant multiple modes.}
    \label{fig:spectrogramh6}
\end{figure*}

In this section, we examine the evolution of spiral arm pattern speeds in the weakly barred, flocculent spiral galaxy of halo 6, which we showed in the previous section exhibits a complex pattern speed in the last gigayear. We focus on four time windows at earlier times than that shown in the right panels of Fig.~\ref{fig:spectrogramall} to tease out this evolution. As in the previous section, each of these time windows span $1.275$ Gyr, which are highlighted in the top row of Fig.~\ref{fig:spectrogramh6}. These time windows are chosen to illustrate the range of spiral pattern speed types in this halo. Fig.~\ref{fig:spectrogramh6} shows the spectrograms for $m=2$ (left column), $m=3$ (middle column), and $m=4$ (right column) for each of these time windows. We show the results for these three modes to get a fuller understanding of the spiral structure, even though $m=2$ is usually the strongest mode. The time window for each spectrogram is indicated in each panel. We discuss the spectrograms for each time window in order of increasing lookback time.

\begin{figure*}
\centering
\includegraphics[scale=0.67,trim={0 0 1.cm 0}, clip]{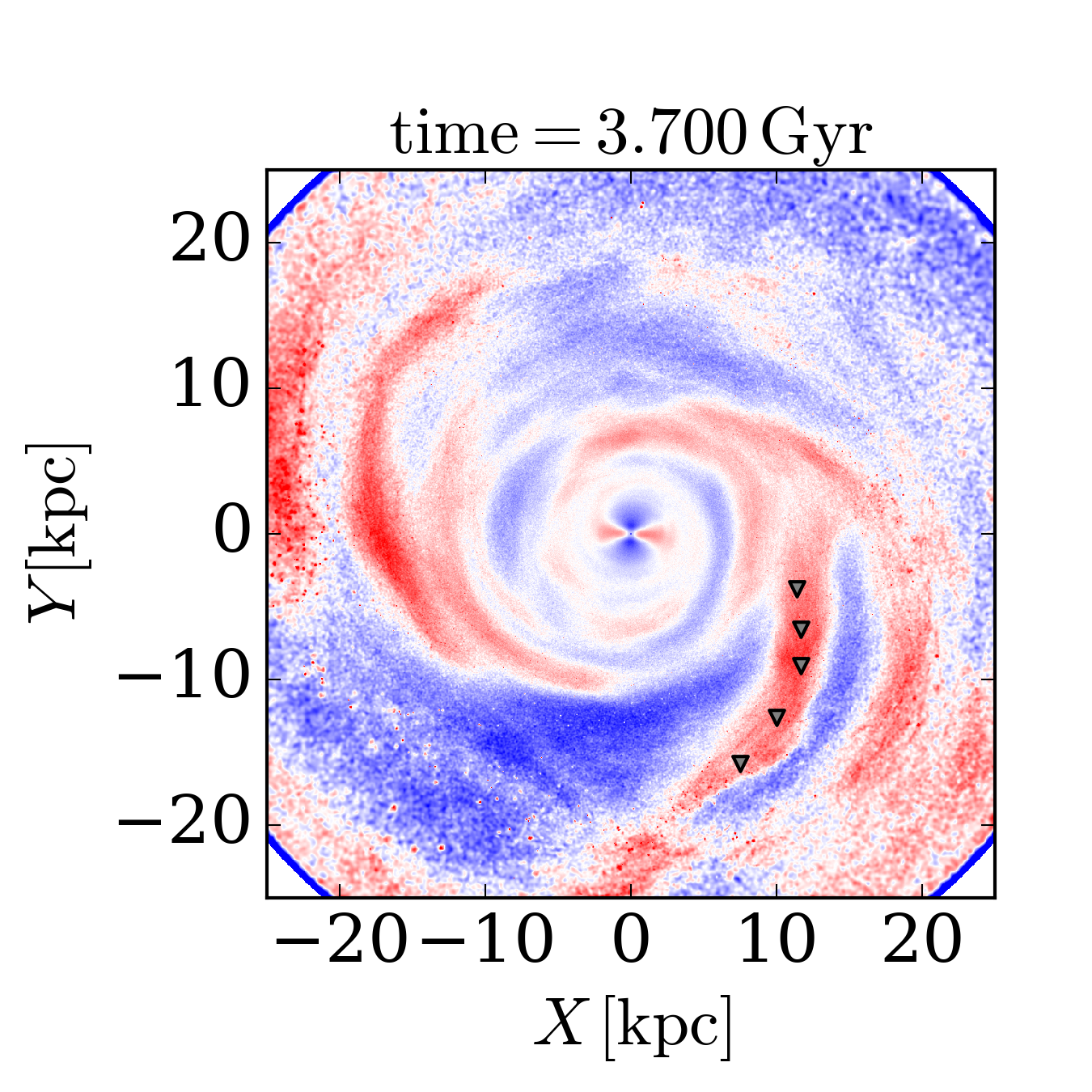}
\includegraphics[scale=0.67,trim={2.7cm 0 1.cm 0}, clip]{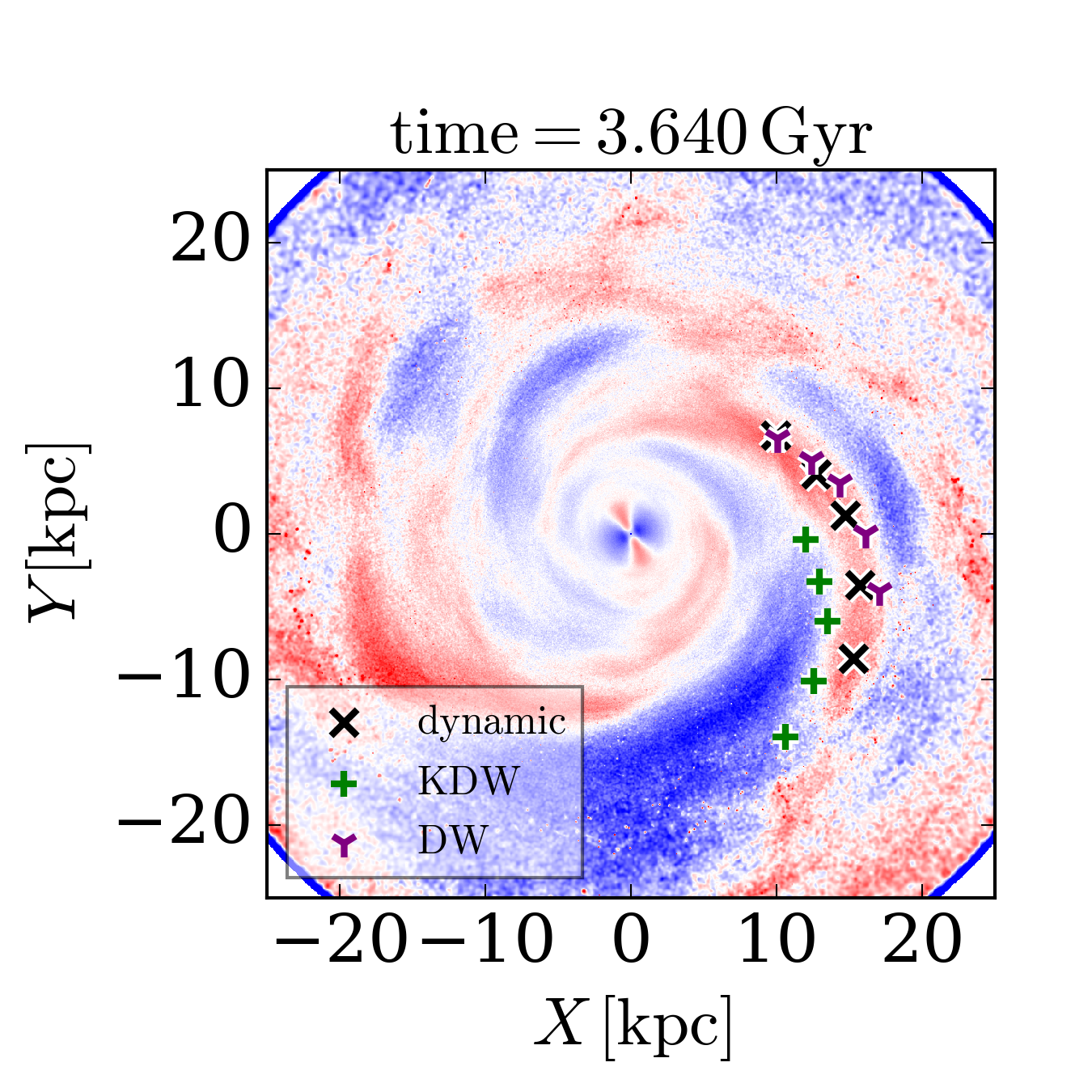}
\includegraphics[scale=0.67,trim={2.7cm 0 1.cm 0}, clip]{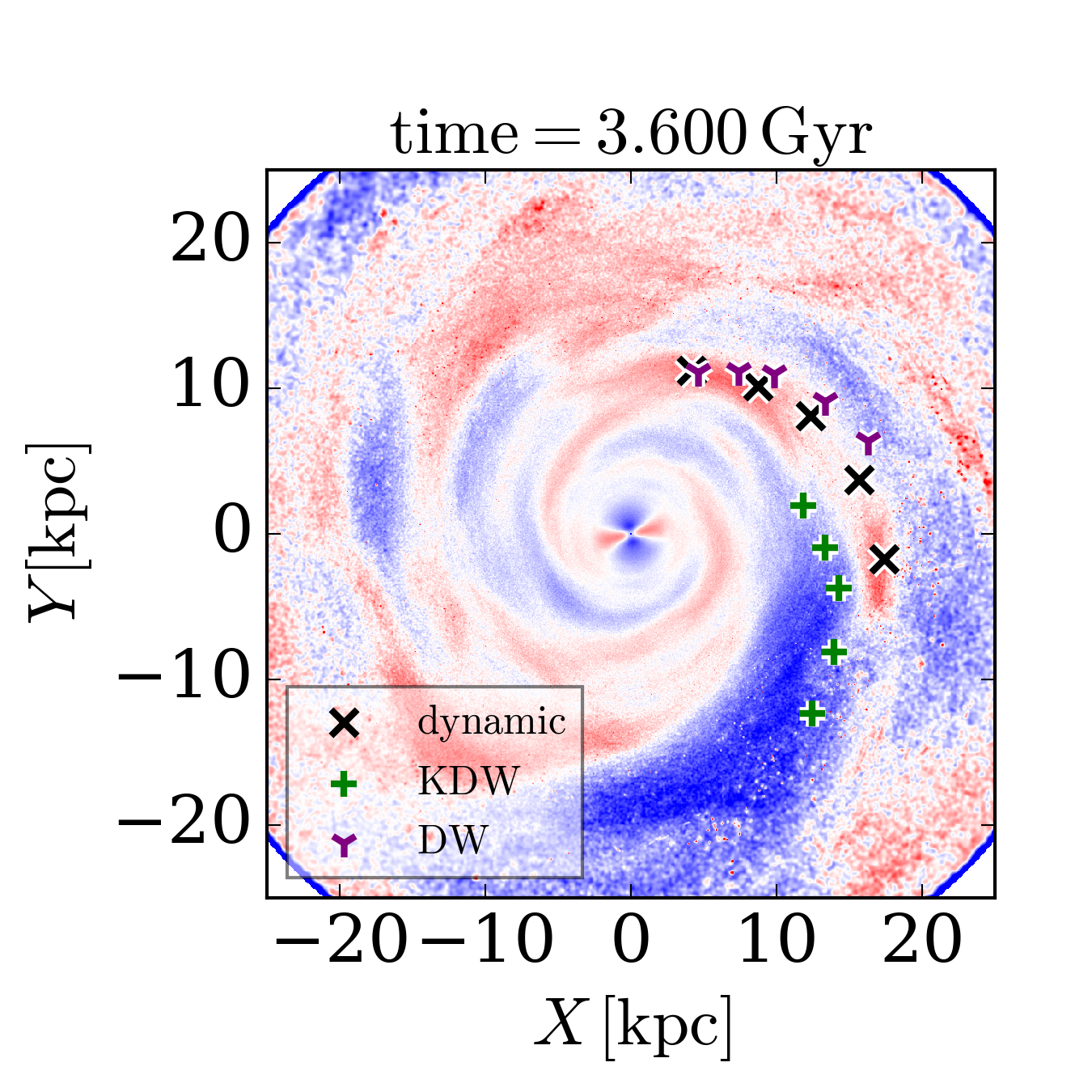}
    \caption{Sequence of stellar surface over-density maps for halo 6 for a series of lookback times in time window C of Fig.~\ref{fig:spectrogramh6}. The left panel shows the locus of a spiral arm  (marked by grey triangle symbols) defined by the peak stellar over-density in azimuth for a series of radial annuli in the outer disc ($10<R<15$ kpc), at $t=3.7$ Gyr. The total lifetime of this arm is of order $\sim 100$ Myr (similar to a dynamical time at these radii) and appears to disrupt at $t=3.6$ (right panel). In the middle and right panels, we show the instantaneous location of the positions shown in the left panel advanced in time according to: the circular velocity curve corresponding to \emph{dynamic} spirals (black crosses); the inner Lindblad resonance curve corresponding to \emph{kinematic density waves} (green plus symbols); and a radially independent angular pattern speed equal to $15$ $\rm km \, s^{-1}$ corresponding to \emph{density waves} (magenta triangles). The dynamic/co-rotating spirals appear to better describe the rotation profile of this spiral arm during this period of time, which is not easily captured in the relatively long baseline of the spectrogram analysis.}
    \label{fig:dynseries}
\end{figure*}

The second row of Fig.~\ref{fig:spectrogramh6} shows the spectrograms for the time window spanning from $1.735$ Gyr to $0.46$ Gyr. It is striking that the spectrogram for the $m=2$ mode shows that most of the power follows the ILR at most radii; the $m=2$ spiral structure behaves like a kinematic density wave, with no trace of the dynamic arm seen in the top-right panel of Fig.~\ref{fig:spectrogramall}. The mean amplitude\footnote{We define the mean amplitude, $A_{\rm sp}$, to be the mean of the $m=2$ Fourier amplitude in the disc region outside of the bar across the time window considered.} of the $m=3$ mode is lower than that of the $m=2$ mode. The contours show that power comparable to the $m=2$ mode exists only in the radial range $18$ kpc - $24$ kpc, which lies close to the circular rotation curve. The top-right panel of Fig.~\ref{fig:spectrogramh6} shows a relatively weak $m=4$ mode with subdominant power everywhere, though the column-normalised power (greyscale) shows aspects of multiple modes.

The third row of Fig.~\ref{fig:spectrogramh6} shows the spectrograms for the time window spanning from $4.205$ Gyr to $2.93$ Gyr. The $m=2$ spectrogram shows a complex pattern speed profile similar to that of the later time window shown in the top-right panel of Fig.~\ref{fig:spectrogramall}: the inner galaxy ($R\rm /kpc < 10$) shows power clustered along the ILR (as expected for kinematic density waves), whereas for the intermediate radii $11<R\rm /kpc <16$, power is clustered mainly around the circular rotation curve, which could be a sign of dynamic/co-rotating spiral arms. This leads also to the rather unexpected inference that these intermediate disc spiral arms rotate faster than the inner disc spiral arms. To further investigate this evolution, we identify a particularly clear spiral arm\footnote{We note that the evolution of this particular spiral arm is representative of other, coexisting spiral arms.} near the middle of this time window, and show a series of face-on over-density maps at three times in Fig.~\ref{fig:dynseries}. In the first panel of Fig.~\ref{fig:dynseries}, we mark the locus of the spiral arm by identifying the azimuth corresponding to the local peak over-density at a series of radii. In the subsequent panels, we mark the prediction for the spiral arm locus according to dynamic (following the rotation curve), kinematic density waves (following the ILR curve), and density waves (a constant pattern speed). This enables us to evaluate the predicted evolution of different theories for individual spiral arms on timescales shorter than the spectrogram bandwidth. We see that the dynamic prediction matches the locus of the spiral arm best: in the middle panel of Fig.~\ref{fig:dynseries}, the kinematic density wave prediction lags too far behind the spiral arm feature, whereas the density wave predicts a too loosely wound shape. The dynamic spiral arm prediction, however, lies on top of the spiral arm. This is even clearer in the right panel of Fig.~\ref{fig:dynseries}, though the spiral arm is already disrupting at this time. We remark also that the spiral arms in the region interior to the marked spiral arm ($5<R\rm /kpc <8$) in this last snapshot rotate more slowly: the innermost cross at $3.64$ Gyr connects to the inner spiral arm, but is seen to have clearly overtaken it at $3.6$ Gyr. An inspection of a movie\footnote{\url{https://wwwmpa.mpa-garching.mpg.de/auriga/movies/Superstars/SpiralArms/stellaroverdensity_halo_6-SF64.mp4}.} showing the evolution of face-on over-density maps with a cadence of 5 Myr confirms that the spiral arms at intermediate radii do indeed rotate faster than the inner disc spirals around this time period.

\begin{figure*}
\centering
\includegraphics[scale=0.67,trim={0 0 1.cm 0}, clip]{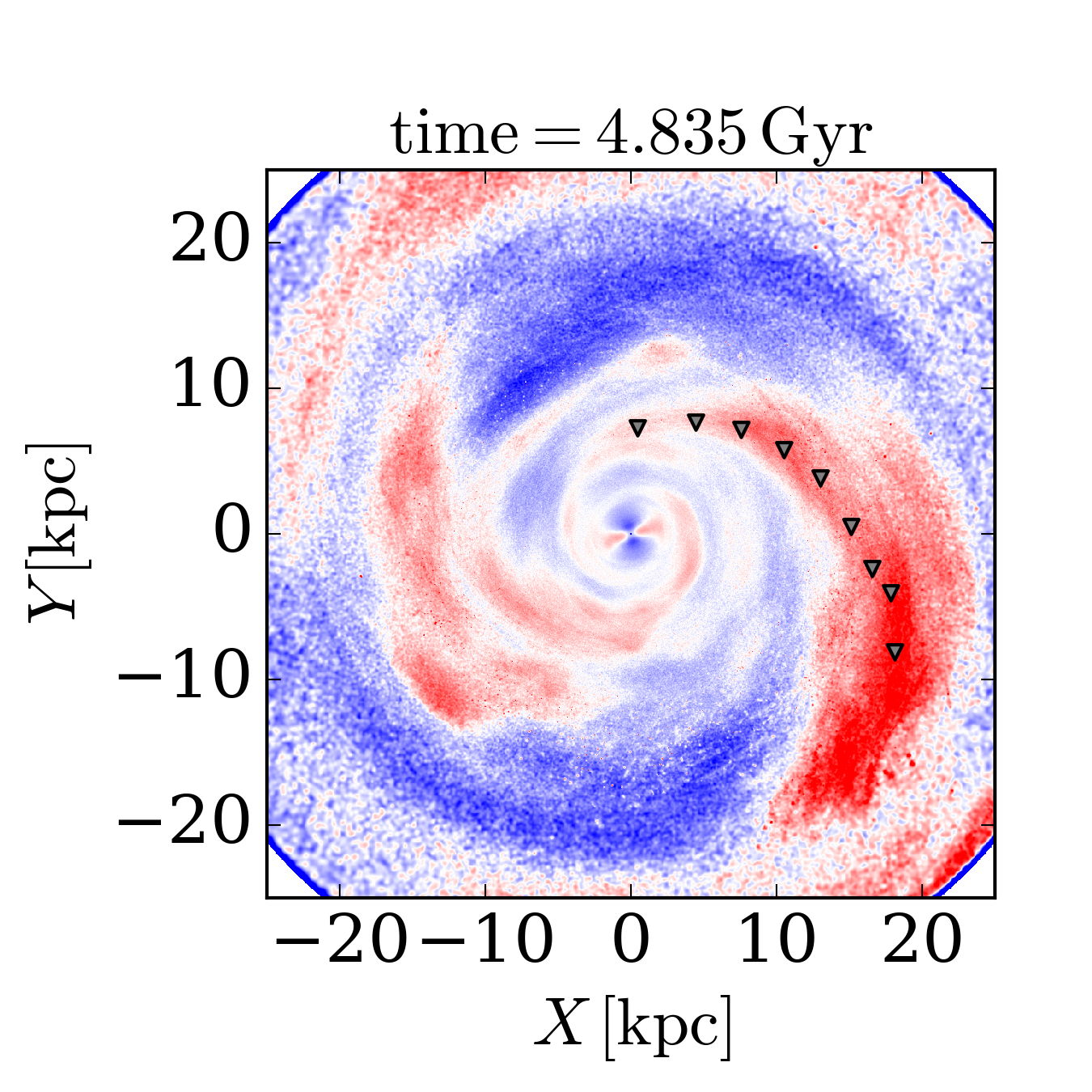}
\includegraphics[scale=0.67,trim={2.7cm 0 1.cm 0}, clip]{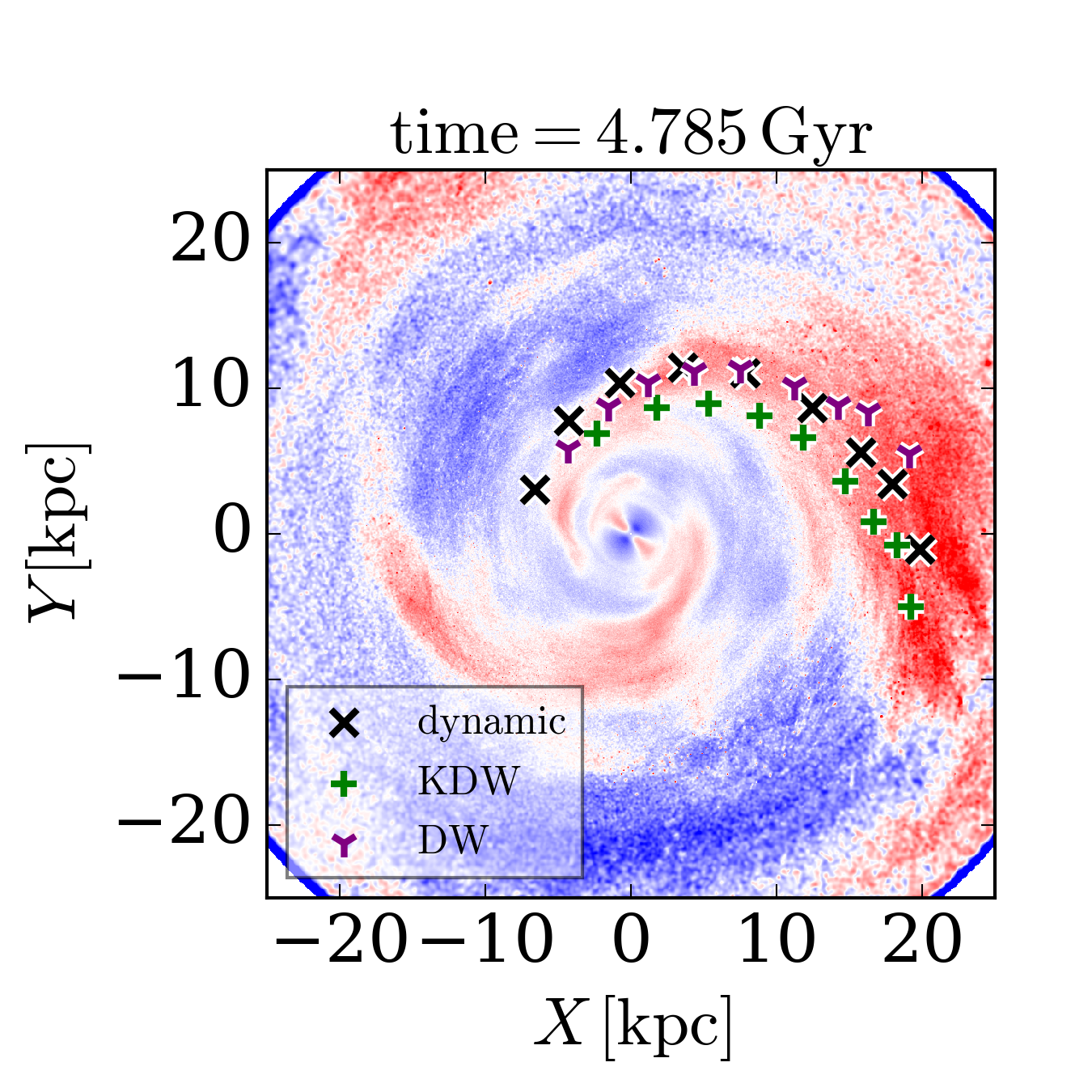}
\includegraphics[scale=0.67,trim={2.7cm 0 1.cm 0}, clip]{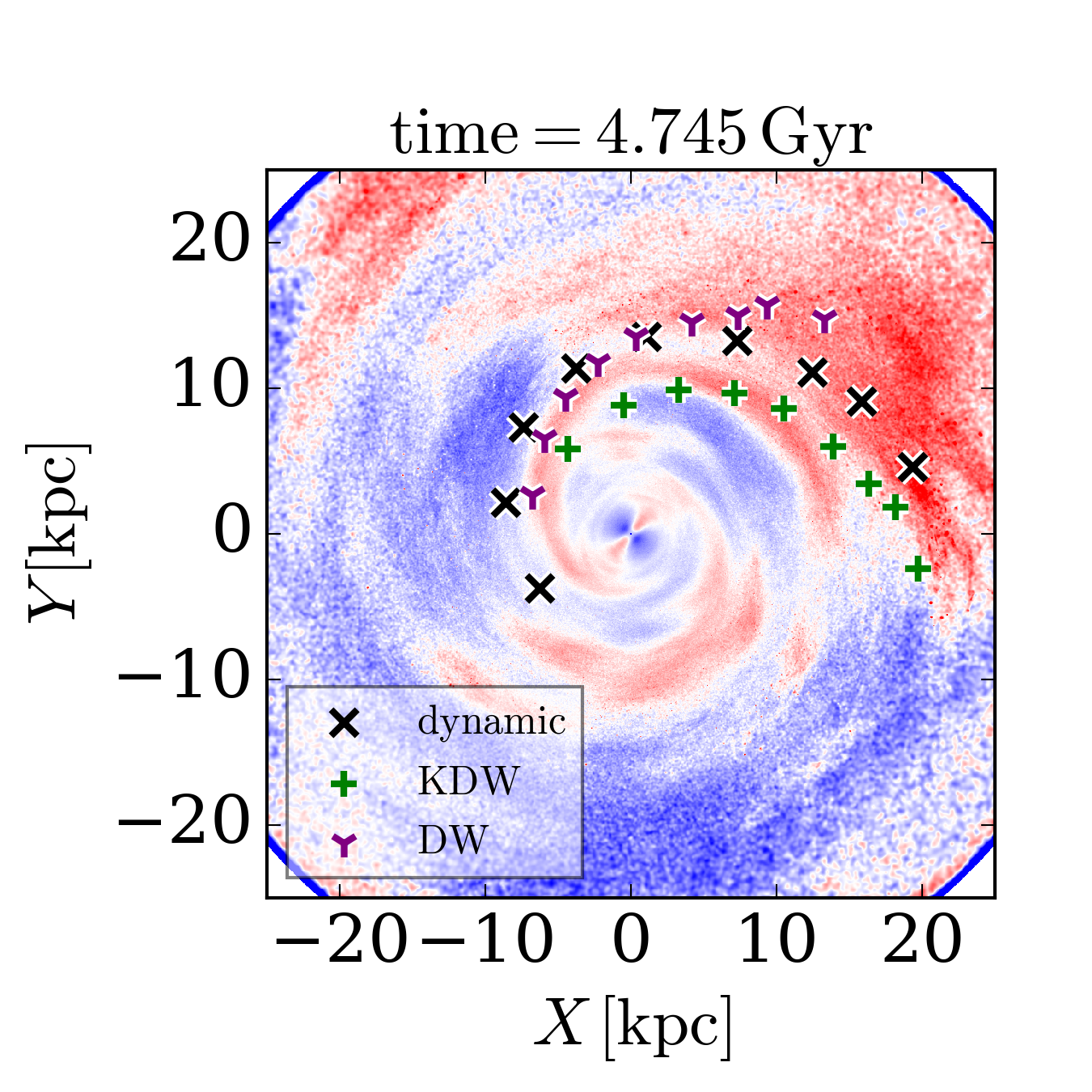}
    \caption{As Fig.~\ref{fig:dynseries}, but showing the evolution of a spiral arm in the time period during which the spectrogram analysis indicates a large-scale density wave (time window B of Fig.~\ref{fig:spectrogramh6}). A density wave with pattern speed $\Omega _p=14$ $\rm km \, s^{-1} \, kpc^{-1}$ tracks the spiral arm consistently until the middle panel. In the third panel, the density wave markers advanced from the first panel track the spiral arm fairly well for radii between 5 kpc and 10 kpc. However, outside of 10 kpc, a separate spiral arm appears to develop, and the density wave markers no longer track a coherent spiral arm locus. This illustrates the complexity and rapid evolution of spiral arms.}
    \label{fig:DWseries}
\end{figure*}

The fourth row of Fig.~\ref{fig:spectrogramh6} shows the spectrograms for the time window spanning from $5.67$ Gyr to $4.395$. In contrast to the later times discussed above, the $m=2$ spectrogram shows that most power belongs to a single pattern speed of $\sim 14$ $\rm km \, s^{-1} \, kpc^{-1}$ that spans the entire allowed range for spiral structure, i.e., from the ILR to the OLR, similar to what is expected for a global density wave. The $m=3$ and $m=4$ spectrograms also highlight power around this frequency (albeit at lower amplitudes). We scrutinised the evolution of the face-on over-density maps within this time window, and we show in Fig.~\ref{fig:DWseries} a series of three snapshots illustrating a clear example of a large-scale spiral arm. The left panel of Fig.~\ref{fig:DWseries} shows the initial locus of a spiral spanning 5 kpc to 20 kpc in radius at a time of $4.835$ Gyr. The second panel shows the spiral 50 Myr later, along with the spiral arm locus predicted by a single density wave with a pattern speed of $\sim 14$ $\rm km \, s^{-1} \, kpc^{-1}$, kinematic density waves, and dynamic (co-rotating) spiral arms. The density wave prediction best matches the location of the spiral arm for radii $\lesssim 10$ kpc; the dynamic (kinematic density wave) markers lie ahead (behind) the spiral arm. At larger radii, the spiral arm structure is azimuthally broad and may comprise more than one spiral feature, which makes it difficult to identify a clear locus. This is seen more clearly in the right panel of Fig.~\ref{fig:DWseries}, in which the spiral splits into at least two separate arms in the outer disc. The clearest of these arms is not part of the original spiral arm identified in the left hand panel: it arises from an amplified over-density patch on the trailing side of the main spiral arm (seen clearly trailing the kinematic density wave markers in the middle panel), and its locus does not align well with any of the predicted locations. 

The large-scale density wave inferred from the $m=2$ spectrogram in the fourth row of Fig.~\ref{fig:spectrogramh6} does not appear to survive longer than $\sim 1$ Gyr; the spectrogram for the time window spanning $6.665$ Gyr to $5.39$ Gyr shown in the fifth row of Fig.~\ref{fig:spectrogramh6} does not show a single dominant frequency, but rather a dominant inner kinematic density wave and what appears to be an outer dynamic spiral. This form of the pattern speed profile is not unlike that shown at later times (third row of Fig.~\ref{fig:spectrogramh6} and top-right panel of Fig.~\ref{fig:spectrogramall}). Therefore, the large scale density wave can only be seen in the time window spanning $5.67$ Gyr to $4.395$ Gyr, and not before or after. It is notable that, during this time window, the galaxy experiences a pericentric passage of a $\sim 10^9$ $\rm M_{\odot}$ satellite. The coincidence of this event with the formation of the large-scale spiral mode indicates the fleeting impact of satellite flybys that can qualitatively change the rotation profile of spiral arms. Interestingly, the $m=3$ spectrogram shown in the middle column of the fifth row of Fig.~\ref{fig:spectrogramh6} shows power concentrated around a pattern speed of around $\sim 22$ $\rm km \, s^{-1} \, kpc^{-1}$ with an amplitude comparable to the $m=2$ mode. The $m=4$ spectrogram in the bottom-right panel shows a decreasing pattern speed for radii larger than 10 kpc, which is qualitatively similar (though weaker in amplitude) to the $m=2$ at these radii. This illustrates the complex and rapidly evolving nature of spiral arms in this simulation.

\section{Discussion}

Our study is the first to link a cosmological spectrum of physical perturbations that arise during galaxy formation and evolution to the emergence and behaviour of spiral structure in Milky Way-mass galaxy simulations with $\sim 10^8$ disc star particles. This number of particles was identified as a threshold to suppress the emergence of artificially seeded spiral arms in $N$-body simulations \citep{DO12}, lending high dynamical fidelity to our results.

The primary finding of this paper is that our cosmological \auriga \superstars simulation suite realises a variety of different spiral arm theories as diagnosed by the radial profiles of spiral arm pattern speeds. We identified two cases in which there is a clear dominant source of perturbation: a strong tidal interaction; and a strong bar, which each produce spiral arms of a fundamentally different nature. For the first time (to our knowledge) in a cosmological hydrodynamic simulation, we found that the strong tidal interaction stimulated the growth of a strong, two-armed grand-design spiral structure which rotated everywhere close to the ILR. This is consistent with kinematic density waves which can be understood as the distortion of stellar orbits onto ovals with differential precession as a function of radius. This pattern would wind up slowly on $\sim$ gigayear timescales. The generation of a kinematic density wave from a strong tidal interaction is consistent with some previous theoretical work on tidally induced spiral structure in which the self-gravity of the disc is relatively weak \citep[e.g.][]{T69,Donner1994}\footnote{Dolfi et al. in prep. shows that the disc of halo 25 has a low central surface density (and is thus weakly self-gravitating) and is also strongly lop-sided as a result of the satellite interaction.}. However, if the lifetime of this spiral is determined by the winding rate, this spiral should survive many galaxy rotations and last several gigayears, which is not necessarily predicted from earlier work. Unfortunately, this interaction occurs at a look back time of $\sim 1$ Gyr, so we are unable to measure its lifetime.

For our other case of spirals in a strongly barred disc (halo 18), we identified, for the first time in a cosmological simulation, a pattern speed profile consistent with manifold spiral theory as described by \citet{RG07} and \citet{Athanassoula2012}. Interestingly, not all of the barred \auriga \superstars simulations show a clear manifold spiral pattern, which suggests that specific conditions may have to be met for manifold spirals to emerge as the dominant spiral structure. In future work, we plan to study the spiral arms in more detail to map out the manifold branches by tracking stellar orbits, understand the onset of their formation, and determine whether the same process (albeit subdominant) occurs in other galaxies. 

Our galaxy that has neither a strong bar nor strong tidal interactions at late times (halo 6) shows a range of pattern speed profiles that evolve on sub-Gyr timescales. In other words, the same galaxy may support different spiral arm types at different times. This result implies that the nature/type of spiral arms (as inferred from the radial profiles of their pattern speeds) are continuously changing as a response to evolving galaxy properties and the many sources of perturbation inherent to galaxy formation, such as the presence of gas (both accreting and within the disc), a non-spherical dark halo, and minor interactions from satellites and dark subhalos. The seemingly rapid transition of spiral pattern speed profile types shown in Fig.~\ref{fig:spectrogramh6} seems to indicate a sensitivity to the aforementioned factors. This provides a complementary view to idealised simulations of isolated discs, in which the gentler secular growth and decay of (multiple) modes is often identified \citep{Sellwood2014,Kumamoto2016}. We have also shown for the first time (to our knowledge) the contemporaneous existence of inner kinematic density waves and intermediate/outer dynamic/co-rotating spiral arms, implying that different  mechanisms are at work in different regions of the disc. This could be related to the radial profiles of stellar (and gas) surface density and radial velocity dispersion - quantities that can play important roles in the formation of spiral arms through processes like swing amplification \citep[e.g.][]{Toomre1964}. We defer a detailed analysis to future work.

The work we have presented is theoretical. To relate our findings and predictions to observations in future studies, we will study observable signatures associated with each type of spiral arm. These include: azimuthal offsets of star-forming gas, stars of different ages, and spiral arm loci, which are expected to be non-zero at specific radii for density wave/multiple mode spirals \citep[see e.g.][]{Querejeta2025}; chemo-kinematic patterns in stars and gas around spiral arms as a result of their dynamical influence, which is expected to vary between different theories \citep{Grand2016,MFS16b,Funakoshi2024}. For now, we focus on comparisons to previous simulation work.

\subsection{Comparison to other cosmological simulations}

Relatively little work has been done on analysing spiral arms in cosmological simulations. \citet{Quinn2025} study the spiral arms in the Milky Way-like spiral galaxies from the FIRE-2 project. In agreement with our work, they find that $m=2$ spirals are the most dominant in their simulation suite. They find that their spiral arms are transient, recurrent features, similar to a wide body of literature as well as the majority of our galaxies (with the exception of the strong bar of halo 18 and strong tidal interaction of halo 25). \citet{Quinn2025} characterise spirals as modes of discrete pattern speed(s), and report a diversity in spiral pattern speeds that evolve with time and sometimes overlap with other modes of different frequencies. In our work, we find also diverse and evolving spiral structure. However, we find also some evidence for qualitatively different types of spiral arms: e.g., manifold spirals and kinematic spirals among others. This may be partially explained by differences in bar strengths between the \auriga and FIRE simulations: \auriga contains a range of bar strengths including strong bars \citep{Fragkoudi2025}, whereas the bars in FIRE are generally relatively weak \citep{Ansar2025}. Therefore, we may not expect that the range of spiral structures found in \auriga and FIRE should match entirely. We also note that the different star formation/feedback models of \auriga and FIRE likely affect the structure and kinematics of the gas and stellar discs, and in turn affect their response to perturbations and the types of spiral structure they can support. 

\citet{Ghosh2025} study the spiral arms in the final ($z=0$) snapshots of the Milky Way/Andromeda analogues of the TNG simulations \citep[e.g.][]{Pillepich2024}. They find that spiral arms are not simply enhancements in recent star formation, in agreement with our findings. They also find that spirals are present in different coeval stellar populations \citep[see also][]{Ardevol2026}, from which they infer that spiral arms are long-lived density waves. The TNG galaxy formation model is similar in many respects to the Auriga model \citep[see][for a description of their main differences]{Grand2024}, therefore we may expect the spiral arms to behave similarly for both physics models. However, we have found a broad range of spiral arm types from our analysis of the dynamical evolution of spiral arms and their pattern speeds. Hence, we speculate that the presence of spiral features in multiple coeval stellar populations may not be an {\it exclusive} prediction of long-lived density waves: indeed, we have verified that the spiral arms in Fig.~\ref{fig:dynseries} appear in both young and old stars, even though they are very short-lived features and their spectrograms do not show density wave-like spirals. Moreover, we show that even the large-scale single pattern speed shown in the third row, left column of Fig.~\ref{fig:spectrogramh6} has a sub-gigayear lifetime. This warrants a future dedicated study of the spiral features in gas and coeval stellar populations for each of the spiral arm types highlighted in our work for a cleaner comparison and understanding of spiral arms in these simulations.

\section{Conclusions}

We have analysed the late-time evolution of spiral arm pattern speeds in the \auriga \superstars cosmological simulations of Milky Way-mass galaxies in the context of current proposed theories of spiral structure. We arrive at the following conclusions:

\begin{itemize}
    \item The simulations show qualitatively diverse spiral arm pattern speed profiles supporting a range of spiral arm theories, including kinematic density waves, dynamic (co-rotating) spirals, manifold spiral arms, and multiple mode density waves. This suggests that the cosmological formation and evolution pathways of disc galaxies determine the nature of the spiral arms that they host, and may go some way to explaining the variety of spiral arm theories supported by observations.
    \item The galaxy with the strongest bar in our \auriga \superstars simulations produces a radial pattern speed profile consistent with manifold spiral arms. Weakly barred galaxies do not appear to produce dominant manifold spirals, suggesting that specific conditions in barred galaxies must be met to produce dominant manifold spirals.
    \item The simulation with the strongest flyby interaction shows the only clear bisymmetric ``grand-design'' type spiral arms, which have a radial pattern speed profile following the inner Lindblad resonance everywhere. Thus, we find that strong tidal interactions can produce large-scale kinematic density waves.
    \item In the absence of a very strong bar and tidal interaction, spiral arms appear to be transient, recurrent features with pattern speeds that continuously change on (sub-)Gigayear timescales. In particular, we find that spiral arms in the same galaxy may transition between kinematic density waves, dynamic spiral arms, and large-scale density waves types on Gigayear timescales. Further, we find that some periods even show separate, contemporaneous patterns  of different nature, specifically kinematic density waves in the inner disc and dynamic spirals in the intermediate/outer disc. 
\end{itemize}

The results presented in this paper mark the first look at spiral arm evolution in the \auriga \superstars simulations. The diversity and rapidly evolving properties of spirals highlighted here provide a foundation on which to build future studies into the origins of spirals, their dependence on galaxy formation physics, and their impacts on galactic structure that can be compared with observations of the Milky Way and nearby galaxies. This incudes the detailed tracking of stellar orbits around spirals, the forward-modelling of velocity and metallicity maps for comparison with IFU observations, and correlating spiral properties with mergers/interactions. To further investigate how bars affect spiral arms, we plan to run a suite of realisations of the same \auriga halos with different random number seeds, as done in \citet{Pakmor2025a}. This will provide a clean comparison of cosmologically simulated spiral galaxies with varying bar strengths (or no bars) that otherwise have the same merger and formation history. Finally, it would be interesting to investigate how galaxy formation physics affects spiral arms and other galaxy properties by, for example, leveraging a neural network-powered exploration of relevant physics model parameter space.

\section*{Acknowledgements}
The authors thank Azadeh Fattahi for producing the initial conditions for the simulations. RJJG thanks Daisuke Kawata and Natsuki Funakoshi for useful simulating discussions.
RJJG acknowledges support from an STFC Ernest Rutherford Fellowship (ST/W003643/1): ``GalaHAD: Galaxy formation with High Accuracy Dynamics''.
RB is supported by the SNSF through the Ambizione Grant PZ00P2\_223532.
FvdV is supported by a Royal Society University Research Fellowship (URF\textbackslash R1\textbackslash191703 and URF\textbackslash R\textbackslash241005).
FAG acknowledges support from the ANID BASAL project FB210003, from the ANID FONDECYT Regular grant 1251493 and from the HORIZON-MSCA-2021-SE-01 Research and Innovation Programme under the Marie Sklodowska-Curie grant agreement number 101086388. FF is supported by a UKRI Future Leaders Fellowship (grant no. MR/X033740/1).
The authors gratefully acknowledge the Gauss Centre for Supercomputing e.V. (www.gauss-centre.eu) for compute time on the GCS Supercomputer
SUPERMUC-NG at Leibniz Supercomputing Centre (www.lrz.de) and the DiRAC@Durham facility which is managed by the Institute for Computational Cosmology on behalf of the STFC DiRAC HPC Facility (www.dirac.ac.uk). The equipment was funded by BEIS capital funding via STFC capital grants ST/K00042X/1, ST/P002293/1, ST/R002371/1 and ST/S002502/1, Durham University and STFC operations grant ST/R000832/1. DiRAC is part of the National e-Infrastructure.

\section*{Data Availability} 
The data underlying this article will be shared on reasonable request to the corresponding author.


\bibliographystyle{mnras}
\bibliography{main.bib}


\appendix

\section{Resolution study}
\label{sec:appendix}

\begin{figure*}
\includegraphics[scale=0.7,trim={0.5cm 0.5cm 2.3cm 0}, clip]{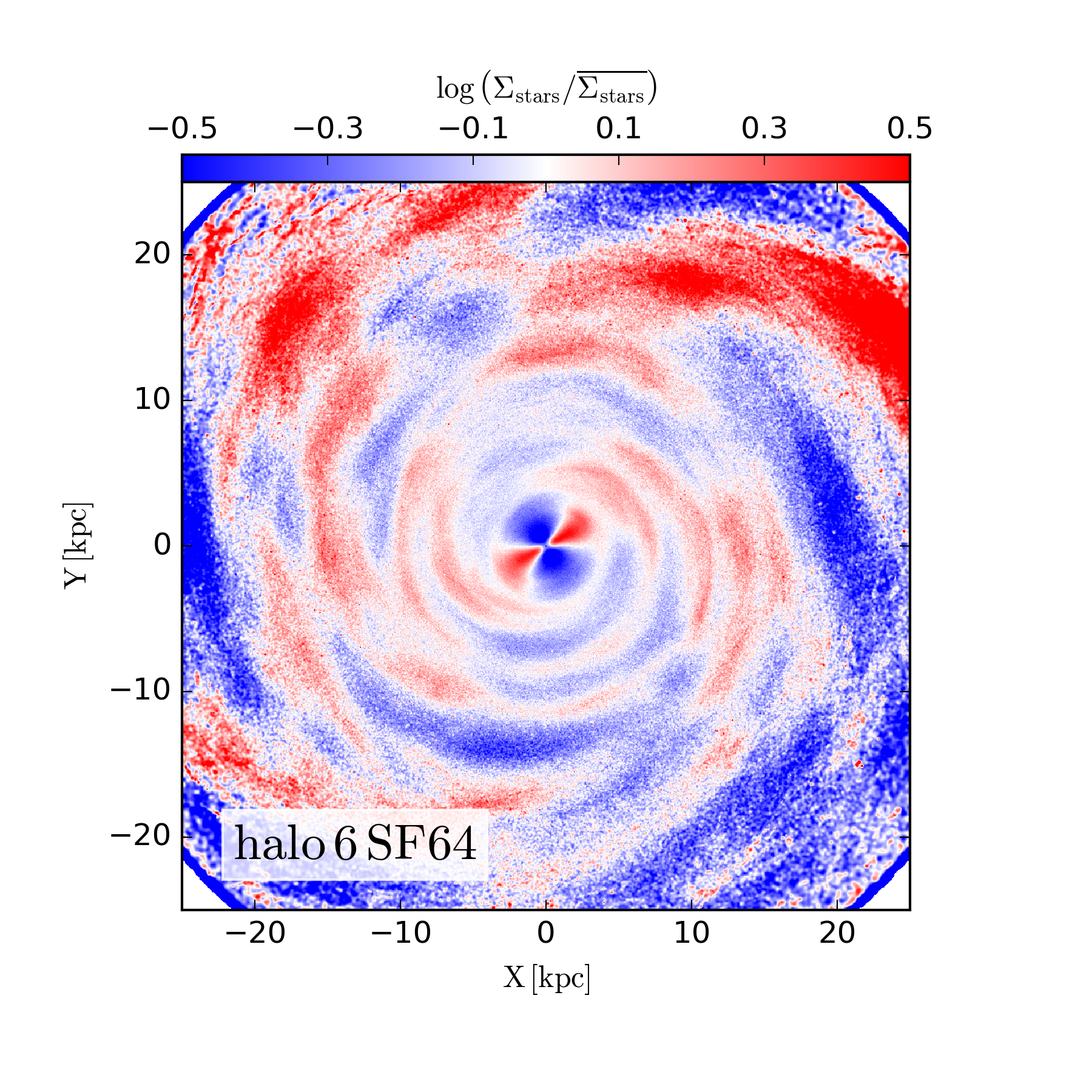}
\includegraphics[scale=0.7,trim={0.5cm 0.5cm 2.3cm 1cm}, clip]{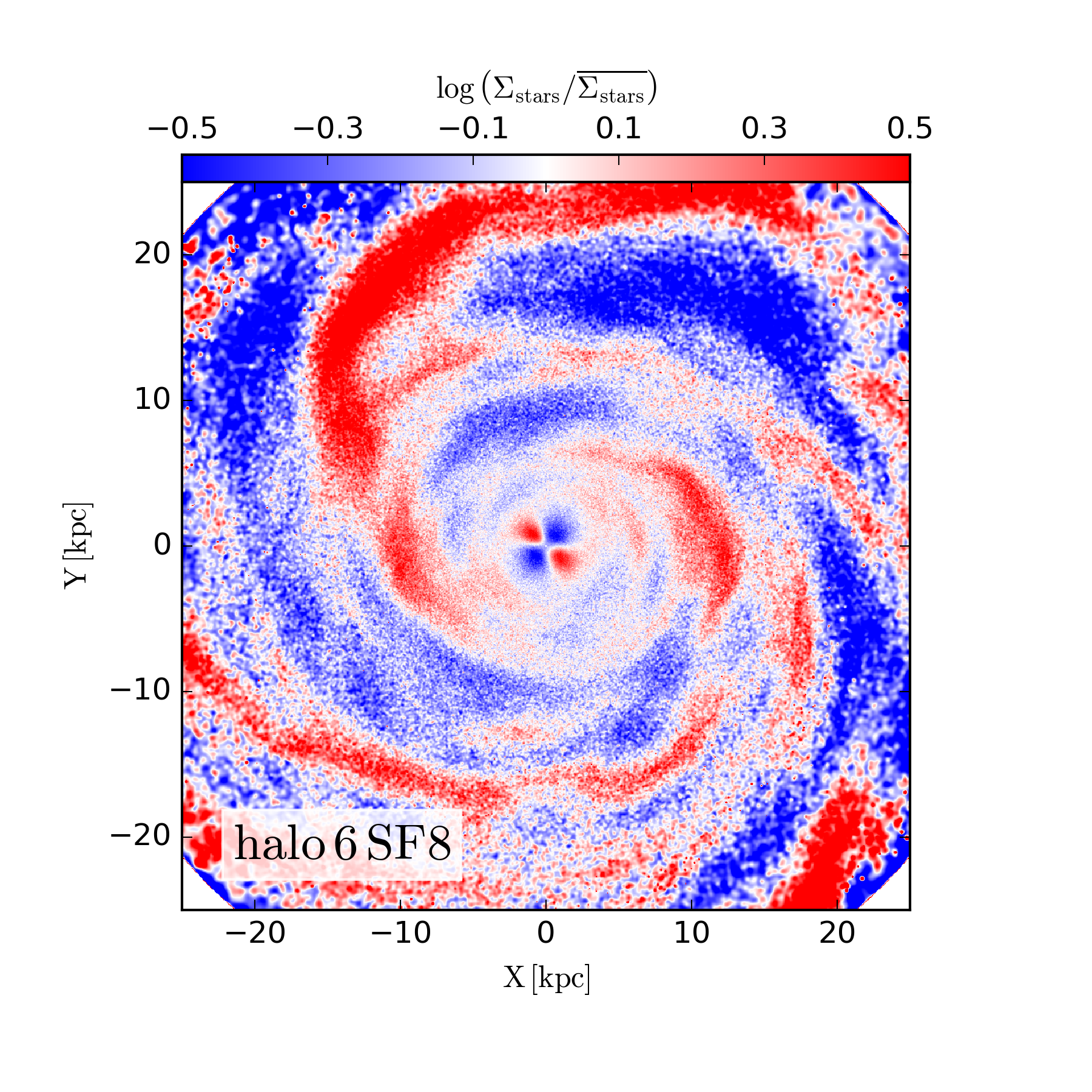}
    \caption{Face-on over-density projections of the final snapshot of the \auriga \superstars version of halo 6 (SF64, left panel) and a version with 8 times lower stellar mass resolution (SF8, right panel).}
    \label{proj_compare}
\end{figure*}

\begin{figure}
\includegraphics[width=\columnwidth,trim={0.5cm 0 2.5cm 1.5cm}, clip]{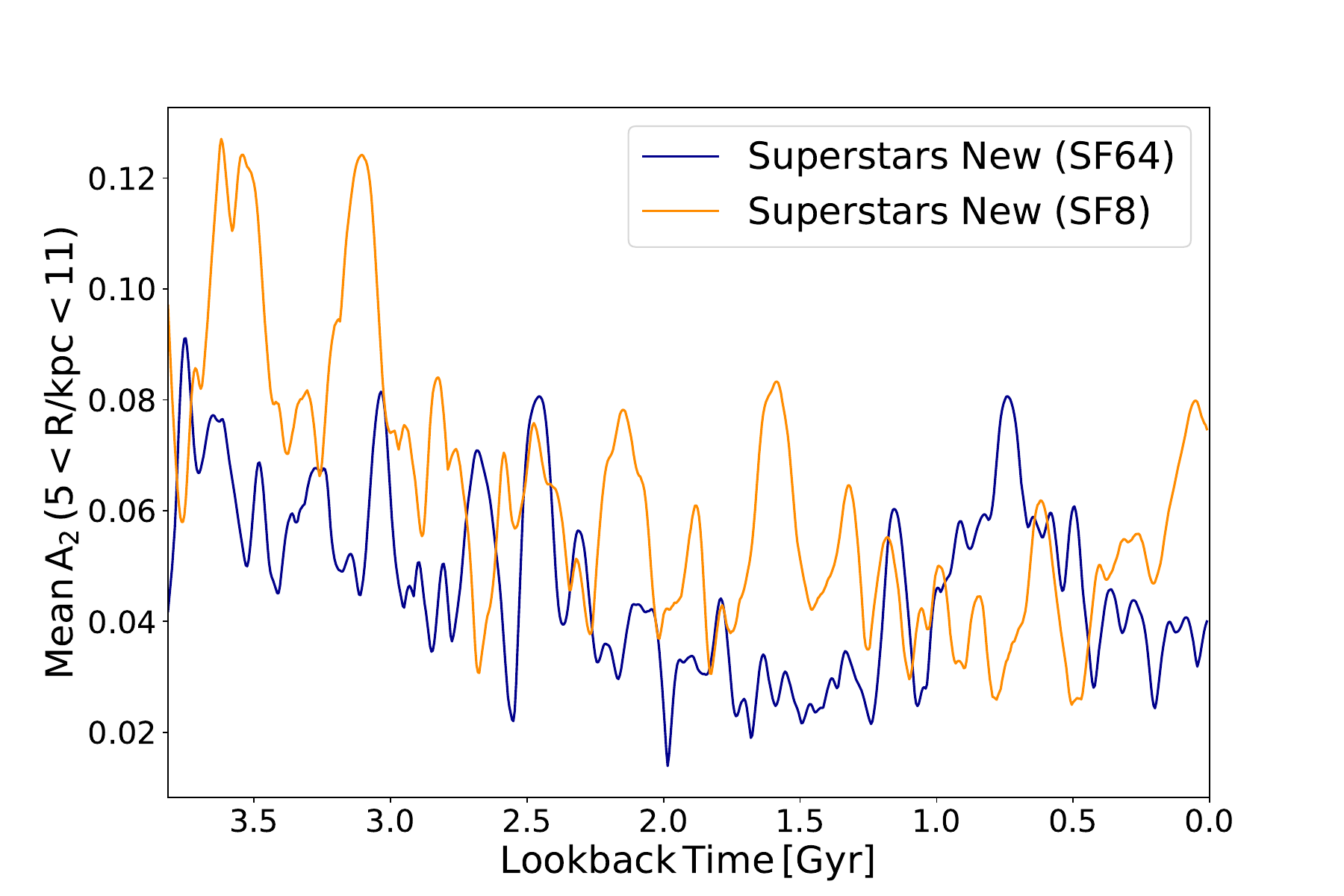}
    \caption{Evolution of the mean $m=2$ Fourier amplitude between the radii of 5 kpc to 11 kpc, for the \auriga \superstars version of halo 6 (SF64) and a version with 8 times lower stellar mass resolution (SF8).}
    \label{amp_compare}
\end{figure}

The left panel of Fig.~\ref{proj_compare} shows the face-on over-density projections of the final snapshot of halo 6 simulated with the superstars method with 64 star particles per star-forming gas cell (SF64, i.e., part of the simulation suite presented in this paper), and the right panel of Fig.~\ref{proj_compare} shows the analogous projection for halo 6 simulated with 8 star particles per star-forming gas cell (SF8). The bar appears marginally longer and stronger for SF64 compared to SF8, whereas the spiral structure appears slightly weaker for SF64 compared to SF8. Nevertheless, the morphology is qualitatively similar. 

In Fig.~\ref{amp_compare}, we show the last 4 Gyr of evolution of the $m=2$ Fourier amplitude of the spiral arms for halo 6 simulated with the superstars method with 64 star particles per star-forming gas cell (SF64) and halo 6 simulated with 8 star particles per star-forming gas cell (SF8). The amplitude is taken to be the average amplitude within the radial range $5<R/\rm{kpc}<11$, which is outside the region of the bar. In each case, the amplitudes are within $0.05$ at all times and show similar fluctuations. In fact, the forms of the curves are so similar that it indicates the timing of the evolution of the SF64 is about $\sim 1$ Gyr behind that of the SF8 run: correcting for this offset makes the amplitudes very similar at most times. Such timing offsets are quite common in re-simulations of galaxies owing to stochastic fluctuations in certain aspects of the physics model \citep[for example, see Fig.~10 of][]{Pakmor2025b}.


\bsp	
\label{lastpage}
\end{document}